# Size-Induced High Electrocaloric Response of Dense Ferroelectric Nanocomposites

Anna N. Morozovska[1*], Oleksandr S. Pylypchuk[1], Serhii Ivanchenko[2], Eugene A. Eliseev[2], Hanna V. Shevliakova[3], Liubomir M. Korolevich[3], Lesya P. Yurchenko[2], Oleksandr V. Shyrokov[2], Nicholas V. Morozovsky[1], Vladimir N. Poroshin[1],

Zdravko Kutnjak[4†], and Victor V. Vainberg[1‡]

[1] Institute of Physics, National Academy of Sciences of Ukraine,

46, pr. Nauky, 03028 Kyiv, Ukraine

[2] Institute for Problems of Materials Science, National Academy of Sciences of Ukraine,

Krjijanovskogo 3, 03142 Kyiv, Ukraine

[3] Department of Microelectronics, National Technical University of Ukraine "Igor Sikorsky Kyiv Polytechnic Institute", pr. Beresteiskyi 37, 03056, Kyiv, Ukraine

[4] Jozef Stefan Institute of Slovenia, Ljubljana, Slovenia

**Abstract**

Analytical results obtained within Landau-Ginzburg-Devonshire approach and effective media models, predict that the synergy of Vegard stresses and size effects can significantly enhance the electrocaloric cooling (up to 7 times) of the $BaTiO_3$ nanoparticles in comparison with a bulk $BaTiO_3$. To compare with the considered effective media models, we measured the capacitance-voltage and current-voltage characteristics of the dense nanocomposites consisting of (28 – 35) vol.% $BaTiO_3$ nanoparticles incorporated in organic polymers and determined experimentally the effective dielectric permittivity and losses of the composites. Generalizing obtained analytical results, various ferroelectric nanoparticles spontaneously stressed by elastic defects, such as oxygen vacancies or any other elastic dipoles, which create a strong chemical pressure, can cause the giant electrocaloric response of dense ferroelectric nanocomposites. We have shown that the advantages of the studied lead-free dense nanocomposites are the good tunability of electrocaloric cooling temperature due to the strain and size effects in ferroelectric nanoparticles and the easy control of the high electrocaloric cooling by electric fields. This makes the dense ferroelectric nanocomposites promising for cooling of conventional and innovative electronic elements, such as FETs with high-temperature superconductor channels.

---

[*] corresponding author, e-mail: anna.n.morozovska@gmail.com
[†] corresponding author, e-mail: zdravko.kutnjak@ijs.si
[‡] corresponding author, e-mail: viktor.vainberg@gmail.com

# I. INTRODUCTION

The electrocaloric (**EC**) effect, consisting of the temperature change under the application or removal of an electric field in adiabatic conditions, and the pyroelectric (**PE**) effect, consisting of the charge or electric field generation under the temperature change, are inherent to ferroelectrics [1], being a consequence of their spontaneous polarization temperature dependence [2, 3], which becomes especially strong in the vicinity of the paraelectric - ferroelectric phase transition [4, 5]. The PE and EC properties determine the indispensable value of ferroelectrics for modern pyroelectric sensors and electrocaloric converters [6, 7, 8].

Polar-active nanocomposites and colloids, which contain ferroelectric nanoparticles of various shapes and sizes, are unique model objects for fundamental studies of the surface, size, and crosstalk effects in the nanoparticles' ensemble [9, 10]; at the same time, being very promising materials for energy storage [11, 12, 13, 14], harvesting [15] and EC applications [16]. Despite the traditional and innovative methods of synthesis, sizes, shape, and polar properties control are well-developed for the uniform [17, 18] and core-shell ferroelectric nanospheres [19, 20], and nanocubes [21, 22], nanorods [23, 24] and nanowires [25, 26], nanoflakes [27] and nanoplates [28, 29], several aspects still contain challenges for preparation technology and uncovering mystery for fundamental theory even for the simplest case of quasi-spherical $BaTiO_3$ nanoparticles of sizes (5 – 50) nm [30, 31, 32].

One of the most comprehensive and general approaches for describing ferroelectric nanoparticles' polar and related properties is the Landau-Ginzburg-Devonshire (**LGD**) phenomenology. LGD combined with the electrostatic equations and finite element modeling (**FEM**) allows establishment of the physical nature of the anomalies in the polar, dielectric, PE, and EC properties of ferroelectric nanoparticles and predicts changes of their phase diagrams with the particle size decreases [33]. The physical nature of the properties' anomalies in the ferroelectric nanoparticles of the size of (5 – 50) nm can be related to the size and surface effects [34, 35, 36], depolarization field originated from the incomplete screening of spontaneous polarization [37, 38], flexoelectricity [39], electrostriction and elastic strains [40, 41, 42].

Many papers are devoted to the theoretical studies of the PE and EC effects in ferroelectric nanoparticles (e.g., [43, 44, 45]). However, the comprehensive theoretical description of the EC response of the dense ferroelectric nanocomposites is still missing. Available theoretical models [46, 47], most of which are applicable only a small volume fraction of the nanoparticles in the composite, use the effective medium approximation for the description of the dielectric and conductive properties. There are many models of dielectric effective media, among which the most known are the Landau approximation of liner mixture [48], Maxwell-Garnett [49] and Bruggeman [50] models for spherical inclusions, and *Lichtenecker-Rother* model of logarithmic mixture [51]. Most models work well for quasi-spherical randomly distributed dielectric nanoparticles in the dielectric media. The applicability of the models is critically sensitive to the crosstalk effects of the polarized nanoparticles [52, 53], and thus, most of them can be invalid for dense nanocomposites, where the volume fraction of ferroelectric nanoparticles is more than 30%. In such a case, it is reasonable to determine the effective dielectric permittivity experimentally and use the obtained value as a material parameter.

Using the LGD theory and effective medium approximation, here we analyze the size-induced and Vegard stress-induced changes of the spontaneous polarization, dielectric permittivity, PE, and EC responses of a spherical BaTiO3 core covered by a charge-screening shell and placed in a dielectric medium. Vegard strains are caused by spontaneously formed elastic defects, such as oxygen vacancies, which concentration is maximal near the nanoparticle surface and monotonically decreases towards its center, creating a diffuse shell with stress and strain gradients. The work has the following structure. The problem formulation for the ferroelectric nanoparticle in the effective dielectric media is described in **Section II.A**, and the approximate analytical description of the quasi-static dependencies of polarization, dielectric, PE, and EC responses is presented in **Section II.B**. Polar and EC properties of a single-domain ferroelectric nanoparticle are analyzed in **Section II.C**. The influence of local fields on the EC response of ferroelectric composites and effective media models are discussed in **Section II.D**. **Section III** is devoted to the experimental determination of effective dielectric permittivity and losses for the dense composites consisting of quasi-spherical BTO nanoparticles incorporated in different organic polymers. The EC response of the ferroelectric composites and some advanced applications of dense lead-free ferroelectric nanocomposites for EC cooling is analyzed in **Sections IV.A** and **IV.B**, respectively. **Section V** contains conclusive remarks. Calculation details and samples characterization are presented in **Supplementary Materials**.

## II. THEORETICAL MODEL

### A. The problem formulation for a ferroelectric nanoparticle in the effective media

Let us consider a spherical BaTiO3 (**BTO**) nanoparticle in the tetragonal phase, whose core of radius $R_c$ is a single domain with a spontaneous polarization $\vec{P}_s$ directed along one of the crystallographic directions (e.g., along the polar axis $x_3$). The nanoparticle geometry is shown in **Fig. 1(a)**. The core permittivity is $\hat{\varepsilon}_c$, which contains a background contribution, $\varepsilon_b$ [54], and the ferroelectric contribution, $\varepsilon_f$. The crystalline and almost insulating core is covered with a "diffuse" charge-screening shell of the average thickness $\Delta R = R_s - R_c$, which is formed spontaneously due to the multiple mechanisms of a spontaneous polarization screening by internal and external charges in nanoscale ferroelectrics (e.g., Ref. [55] and refs therein). The shell is semiconducting due to the high concentration of free charges and strained due to the high concentration of elastic defects. The free charges provide effective screening of the core spontaneous polarization and prevent domain formation. The effective screening length in the shell, $\lambda$, is very small (less than 1 nm), and its relative dielectric permittivity tensor, $\varepsilon_{ij}^S$, is isotropic, $\varepsilon_{ij}^S = \delta_{ij}\varepsilon_s(\vec{r})$, and can be coordinate–dependent as anticipated in the inhomogeneous paraelectric shell region. The elastic defects can induce strong Vegard strains and stresses in the diffuse shell [56, 57]. The Vegard strain tensor, denoted as $w_{ij}^S(\vec{r})$, is regarded as isotropic, $w_{ij}^S = \delta_{ij}w_s$, where $\delta_{ij}$ is the Kronecker-delta symbol and $w_s$ is the tensor magnitude. As a rule, $w_s$ cannot exceed $(1 - 3)\%$, and thus corresponding Vegard stresses, $\sigma_{ij}^S(\vec{r}) = c_{ijkl}w_{ij}^S(\vec{r})$, cannot exceed $(2 - 6)$ GPa because elastic stiffnesses $|c_{ijkl}| \geq 2 \cdot 10^{11}$ Pa. The nanoparticles are placed in a polymer matrix with the dielectric permittivity $\varepsilon_M$. The dielectric properties of the nanocomposite system "ferroelectric nanoparticles + polymer" are considered in the effective media approximation, where the dependence of the effective media dielectric permittivity, $\varepsilon_{eff}$, on $\varepsilon_M$, $\varepsilon_s$, $\varepsilon_c$, and nanoparticles volume fraction $v_P$ should be determined in a self-consistent way.

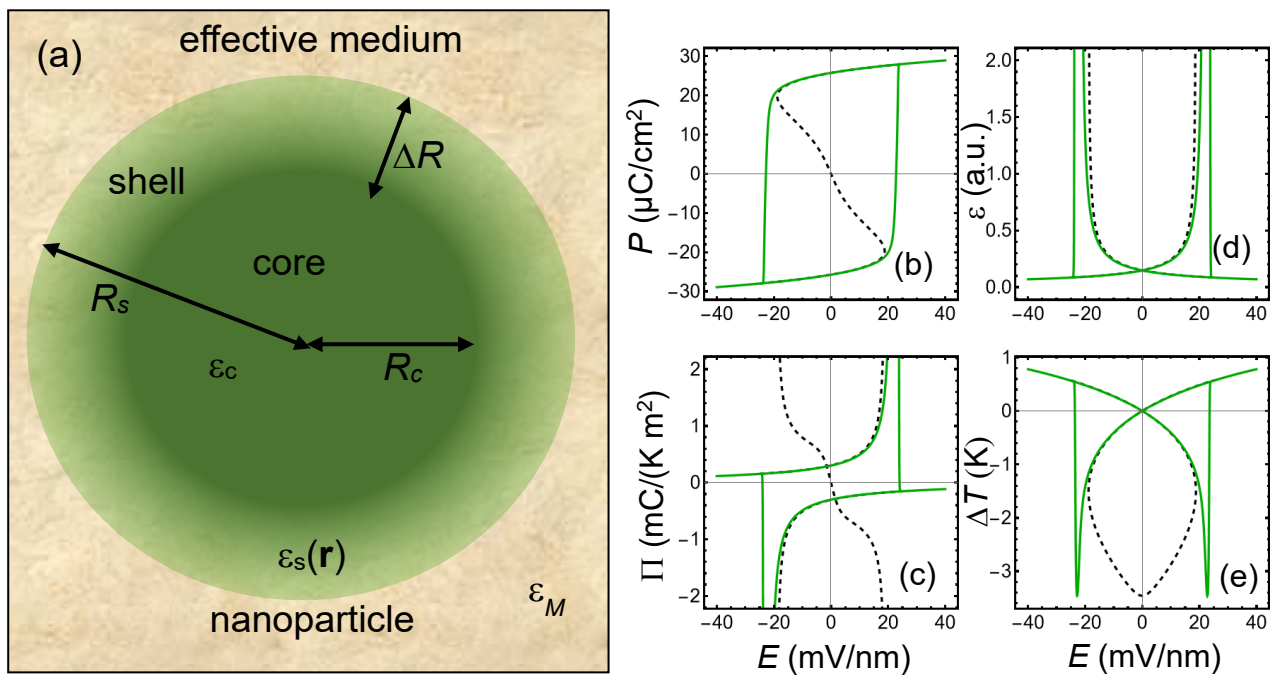

**FIGURE 1.** (a) A nanoparticle geometry: a spherical ferroelectric core of radius $R_c$ is covered with a "diffuse" shell of thickness $\Delta R$ and outer radius $R_s$. The nanoparticle is placed in an isotropic dielectric medium; $\varepsilon_c$, $\varepsilon_s$, and $\varepsilon_M$ are the core, shell, and surrounding media dielectric permittivities. Typical quasi-static hysteresis loops (green solid curves) and static dependencies (black dashed curves) of the polarization $P$ (b), PE coefficient $\Pi$ (c), relative dielectric permittivity $\varepsilon_c$ (d), and the EC temperature change $\Delta T_{EC}$ (e) of a relatively small stressed BTO nanoparticle ($R_s = 12$ nm, $\Delta R = 5$ nm), Vegard stress $\sigma_s = 3$ GPa and temperature $T = 200$ K.

Our goal is to calculate the low-frequency (i.e., quasi-static) and static field dependencies of the polarization, PE coefficient, relative dielectric permittivity, and the EC temperature change of a single-domain nanoparticle, which are schematically shown in **Fig. 1(b)**, **1(c)**, **1(d)** and **1(e)**, by green solid and black dashed curves, respectively. To reach the goal, we use the LGD phenomenological approach.

The LGD free energy density includes the Landau-Devonshire expansion in even powers of the polarization $P_3$ (up to the 8-th power), the Ginzburg gradient energy and the elastic, electrostriction, and flexoelectric energies, which are listed in **Appendix A1** of **Supplementary Materials** [58].

The dynamics of quasi-static polarization $P_3$, PE coefficient $\Pi_3$ and dielectric susceptibility $\chi_{33}$ in the external field $E_3$ follows from the time-dependent LGD equations, which are listed in **Appendix A2**. The value $E_3$ is an electric field component, which is a superposition of external and depolarization fields, $E_3^0$ and $E_3^e$, respectively. The quasi-static field $E_3$ is related to the electric potential $\varphi$ as $E_3 = -\frac{\partial \varphi}{\partial x_3}$. The potential $\varphi$ satisfies the Poisson equation inside the nanoparticle and the Laplace equation outside the charge-screening shell (see **Appendix A1** for details).

The LGD parameters of a bulk BTO used in our calculations are taken from Pertsev et al. [59] and Wang et al.[60]. They are listed in **Table AI** in **Appendix A1**.

### B. Analytical description of the quasi-static polar, dielectric and electrocaloric properties of a single-domain ferroelectric nanoparticle

In the case of natural boundary conditions used here, small screening length $\lambda \leq 0.5$ nm and relatively high gradient coefficients of polarization, $|g_{ijkl}| > 10^{-11}$ C$^{-2}$m$^3$J, polarization gradient effects can be neglected in a single-domain state, which is revealed to have minimal energy compared to polydomain states. The field dependence of a static single-domain polarization can be found from the following equation [32]:

$$\alpha^* P_3 + \beta^* P_3^3 + \gamma P_3^5 + \delta P_3^7 = E_3^e \qquad (1)$$

The parameters $\alpha = \alpha_T(T - T_C^\alpha)$, $\beta = \beta_0 + \beta_T(T - T_C^\beta)$, $\gamma = \gamma_0 + \gamma_T(T - T_C^\gamma)$, and $\delta$ are the 2-nd, 4-th, 6-th and 8-th order Landau expansion coefficients in the $P_3$-powers of the free energy corresponding to the bulk BTO. The depolarization field, $E_3^d$, and stresses, $\sigma_i$, contribute to the "renormalization" of the coefficient $\alpha(T)$, which becomes the temperature-, radius-, stress-, and screening length-dependent function $\alpha^*$ [61]:

$$\alpha^*(T, R_c, \vec{r}) = \alpha(T) + \frac{1}{\varepsilon_0(\varepsilon_b + 2\varepsilon_s + R_c/\lambda)} - 2Q_{i3}\sigma_i(\vec{r}) - W_{ij3}\sigma_i(\vec{r})\sigma_j(\vec{r}). \qquad (2)$$

The derivation of the depolarization field contribution, $\frac{1}{\varepsilon_0(\varepsilon_b+2\varepsilon_s+R_c/\lambda)}$, in Eq.(3b) is given in Ref. [62], $\varepsilon_0 = 8.85$ pF/m and $\varepsilon_b = 7$. Due to the nonlinear electrostriction coupling, the coefficient $\beta^*$ is "renormalized" by elastic stresses as

$$\beta^*(\vec{r}) = \beta - 4Z_{i33}\sigma_i(\vec{r}). \tag{3}$$

The values $Q_{i3}$, $Z_{i33}$, and $W_{ij3}$ are the components of a linear and two nonlinear electrostriction strain tensors in the Voigt notation, respectively [63].

In the right-hand side of Eq.(1) $E_3^e$ is the static external field inside the core, for which the estimate $E_3^e \approx \frac{3\varepsilon_s E_3^0}{\varepsilon_b+2\varepsilon_s+R_c/\lambda}$ is valid. Only if $\lambda(E_3^0) \gg R_c$ and $\varepsilon_s \approx \varepsilon_{eff}$ the external field in the core is in the same order as the applied field $E_3^0$, otherwise $|E_3^e| \ll |E_3^0|$.

Elastic stresses in the core, $\sigma_i$, induced by the Vegard strains in the shell, can be calculated analytically using the method of successive approximations [32]. For a spherical nanoparticle, only the following combinations of linear and nonlinear electrostriction coupling constants $Q_{ij}$, $Z_{ijk}$ and $W_{ijk}$ are included in the solution [32]:

$$Q_c = Q_{11} + 2Q_{12}, \quad Z_c = Z_{111} + 2Z_{211}, \quad W_c = W_{111} + 2W_{112} + 2W_{123} + 4W_{122}. \tag{4}$$

For numerical calculations, we use the combinations of the following values, $Q_c = 0.0067$ C$^{-2}\cdot$m$^4$, $Z_c = 0.25$ C$^{-4}\cdot$m$^8$ and $0 \leq W_c \leq 10^{-12}$ C$^{-2}\cdot$m$^4$ Pa$^{-1}$, which were experimentally determined from the piezoelectric coefficients and spontaneous strains of BTO at room temperature [64].

The average Vegard stresses induced by the Vegard strains have the form:

$$\langle\sigma_{11}^S\rangle = \langle\sigma_{22}^S\rangle = \langle\sigma_{33}^S\rangle = \frac{R_s^3 - R_c^3}{R_s^3}\langle\sigma_s\rangle, \qquad \langle\sigma_s\rangle = \frac{\langle w_s\rangle}{s_{11}+s_{12}}. \tag{5}$$

Here $\langle\sigma_s\rangle$ and $\langle w_s\rangle$ are the average amplitudes of the Vegard stress and strain in the diffuse shell, respectively.

After substitution of the Vegard stresses (5) into Eqs.(2) and (3), and elementary transformations, the following equation for the spontaneous polarization with renormalized coefficients is obtained:

$$\alpha_R P_3 + \beta_R P_3^3 + \gamma P_3^5 + \delta P_3^7 = E_3^e. \tag{6}$$

The renormalized coefficients are given by expressions:

$$\alpha_R = \alpha + \frac{1}{\varepsilon_0(\varepsilon_b+2\langle\varepsilon_s\rangle+R_c/\lambda)} - 2Q_c\frac{R_s^3 - R_c^3}{R_s^3}\langle\sigma_s\rangle - W_c\left(\frac{R_s^3 - R_c^3}{R_s^3}\langle\sigma_s\rangle\right)^2, \tag{7a}$$

$$\beta_R = \beta - 4Z_c\frac{R_s^3 - R_c^3}{R_s^3}\langle\sigma_s\rangle. \tag{7b}$$

The renormalization of the coefficients in Eq.(7) is proportional to the ratio $\frac{R_s^3 - R_c^3}{3R_s^3}$, which is close to $\frac{\Delta R}{R_s}$ for thin shells with $\Delta R \ll R_c$. The renormalization is the most significant for small nanoparticles with $\Delta R \sim R_c$; it vanishes for $R_c \to R_s$ and is absent in thin films, where the curvature disappears, $R_c$ and $R_s$ tend to infinity, and their finite difference $\Delta R$ becomes the film thickness. Also, the condition $W_c \geq 0$ should be valid for the stability of the Eq.(6) solutions.

The field dependence of a static single-domain PE coefficient $\Pi_3$, dielectric susceptibility $\chi_{33}$, and the EC temperature change $\Delta T_{EC}$ in an external field $E_3^e$ can be found from the following expressions [32, 33]:

$$\Pi_3(E_3^e) = -\left(\frac{\partial P_3}{\partial T}\right)_{E_3^e}, \quad \chi_{33}(E_3^e) = \frac{\partial P_3}{\partial E_3^e}, \tag{8}$$

$$\Delta T_{EC}(E_3^e) \cong T \int_0^{E_3^e} \frac{1}{\rho_P C_P} \Pi_3 \, dE \approx \frac{T}{\rho C_P} \left(\frac{\alpha_T}{2}[P_3^2(E_3^e) - P_3^2(0)] + \frac{\beta_T}{4}[P_3^4(E_3^e) - P_3^4(0)] + \frac{\gamma_T}{6}[P_3^6(E_3^e) - P_3^6(0)]\right), \tag{9}$$

where $\alpha_T = \frac{\partial \alpha_R}{\partial T}$, $\beta_T = \frac{\partial \beta_R}{\partial T}$ and $\gamma_T = \frac{\partial \gamma}{\partial T}$.

### C. Stress-induced enhancement of the electrocaloric properties of a single-domain ferroelectric nanoparticle

Analytical and numerical results presented in the section are obtained and visualized in Mathematica 13.2 [65]. The calculations were performed for a single-domain BTO nanoparticle with a small screening length ($\lambda = 0.1 - 0.5$ nm), small shell thickness ($\Delta R = 1 - 5$ nm), high average dielectric permittivity of the diffuse shell ($\langle \varepsilon_s \rangle \approx 100 - 300$), wide range of surrounding media permittivity ($\varepsilon_M = 3 - 30$), and temperature range (0 – 400) K. For the parameters, the size effects in Eqs.(6) are very significant for $R_c < 100$ nm and become negligible for $R_c > 500$ nm.

In dependence on the values of $\lambda$ and $\langle \varepsilon_s \rangle$, small nanoparticles are paraelectric due to the depolarization field effect, whose influence is very strong, being proportional to $\frac{\lambda}{R_c}$ in accordance with the term $\frac{1}{\varepsilon_0(\varepsilon_b + 2\langle \varepsilon_s \rangle + R_c/\lambda)}$ in Eq.(7a). For typical parameters $\lambda = 0.5$ nm, $\Delta R = 5$ nm and $\varepsilon_s = 150$, used in **Fig. 2**, we consider the actual range of core radii, $6 < R_c < 60$ nm, for which the nanoparticle core can be ferroelectric, and the role of Vegard stresses is pronounced.

The spontaneous polarization, $P_s$, PE coefficient, $\Pi_s$, linear dielectric permittivity in the polar direction, $\varepsilon_f = \varepsilon_{33}$, and EC temperature change, $\Delta T_{EC}$, as a function of temperature $T$ and Vegard stresses $\sigma_s$, are shown in **Figs. 2** for core radii $R_c = 6$ nm, 12 nm, and 60 nm, respectively. The abbreviations "PEP" and "FEP" mean the paraelectric and ferroelectric phases, respectively. $P_s$, $\Pi_s$ and $\varepsilon_f$ are calculated from Eq.(8) for $E_3^e \to 0$, and $\Delta T_{EC}$ is calculated from Eq.(9) for $E_3^e \to E_c$, where $E_c$ is the coercive field. Thin white curves in the plots of $\varepsilon_f$ and $\Pi_s$ correspond to the regions where these values diverge at the boundary of the paraelectric-ferroelectric phase transition.

Since $Q_c > 0$ for BTO, compressive Vegard stresses $\sigma_s < 0$ suppress $P_s$ and can induce the paraelectric state. In contrast, tensile Vegard stresses $\sigma_s > 0$ increase $P_s$ and support the ferroelectric state in the BTO core [see **Figs. 2(a)-(c)**]. Due to the synergy of size and Vegard effects, the core spontaneous polarization $P_s$ reaches 55 µC/cm² (for $R_c = 6$ nm and $\sigma_s = 6$ GPa) and 45 µC/cm² (for $R_c = 60$ nm and $\sigma_s = 6$ GPa) in the temperature range (298 – 388) K, which is significantly higher than the values (26 – 21) µC/cm² corresponding to the unstressed bulk BTO.

The $\Pi_s$ and $\varepsilon_f$ diverge near the boundary of the stress-induced paraelectric-ferroelectric phase transition, and the divergence is shown by thin white lines in **Figs. 2(d-f)** and **2(j-i)**. Despite the $\Pi_s$ does not exceed (1 – 2) mC/(K m²) far from the paraelectric-ferroelectric boundary, the field behavior of $\Pi_s$ determines the features of EC properties in accordance with Eq. (7b).

The EC cooling temperature $\Delta T_{EC}$ can be more than -17 K (for $R_c = 6$ nm and $\sigma_s = 6$ GPa) and about -6 K (for $R_c = 60$ nm and $\sigma_s = 6$ GPa) far from the boundary of the paraelectric-ferroelectric phase transition [see **Figs. 2(j)-(l)**]. Note that $\Delta T_{EC}$ cannot exceed -2.5 K neither for a bulk BTO nor for unstrained BTO nanoparticles in accordance with earlier LGD-based theoretical calculations [33]. The negative sign of the EC effect and its maximum, which appear with an increase in the spontaneous polarization, is related to the features of the LGD free energy for BTO, where all the expansion coefficients from the 2-nd to the 6-th power of polarization depend on temperature. According to Eq.(9), this leads to several contributions to the EC effect, proportional to the even powers of $P_s$ and to the factor $\frac{T}{\rho C_P}$:

$$\Delta T_{EC}(E_c) \approx -\frac{T}{\rho C_P}\left(\frac{\alpha_T}{2}P_s^2 + \frac{\beta_T}{4}P_s^4 + \frac{\gamma_T}{6}P_s^6\right), \quad \Delta Q_{EC}(E_c) \approx -T\left(\frac{\alpha_T}{2}P_s^2 + \frac{\beta_T}{4}P_s^4 + \frac{\gamma_T}{6}P_s^6\right). \quad (10)$$

where we used that $P_3(E_c) \to 0$ in the case of quasi-static and dynamic polarization reversal.

Hence, the synergy of size effects and Vegard strains can significantly improve the spontaneous polarization (1.5 – 2 times) and the enhance EC properties (2 – 7 times) of a stressed BTO nanoparticle in comparison with a bulk BTO and unstrained BTO nanoparticles. The conclusion can be general for other ferroelectric nanoparticles spontaneously stressed by inherent (or incorporated) elastic defects, such as oxygen vacancies or any other elastic dipoles, which create enough strong chemical pressure in the particles and cause their giant EC response.

The advantages of the BTO nanoparticles are good tunability of their EC cooling temperature due to the stresses and size effects in the nanoparticles (see green and blue areas in **Fig. 2(j)-(l)** for the EC response corresponding to the wide temperature range 50 – 300 K).

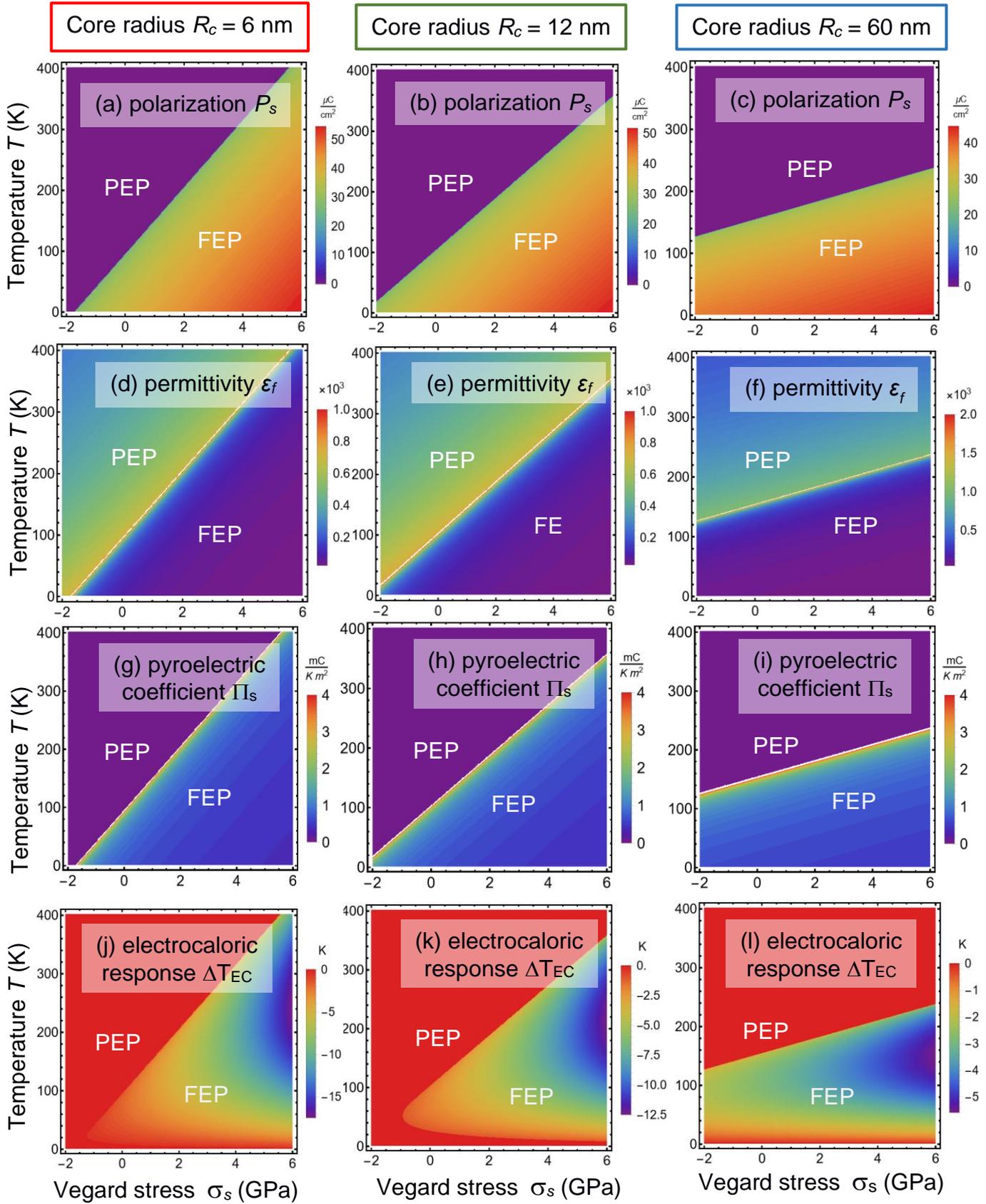

**FIGURE 2.** The spontaneous polarization $P_s$ (**a-c**), relative dielectric permittivity $\varepsilon_f$ (**d-f**), PE coefficient $\Pi_s$ (**g-i**), and the EC temperature change $\Delta T_{EC}$ (**j-l**) of a single-domain BTO nanoparticle as a function of temperature $T$ and Vegard stresses $\langle \sigma_s \rangle$, calculated for the core radius $R_c = 6$ nm (the left column), 12 nm (the middle column), and 60 nm (the right column), $\lambda = 0.5$ nm, $\Delta R = 5$ nm and $\langle \varepsilon_s \rangle = 150$. The color scale

shows the values of $P_s$ in μC/cm², $\Pi_s$ in mC/(K m²), $\varepsilon_f$ in dimensionless units, and $\Delta T_{EC}$ in K, respectively. Thin white areas in the plots of $\varepsilon_f$ and $\Pi_s$ correspond to the regions where these values diverge. The abbreviations "PEP" and "FEP" mean the paraelectric and ferroelectric phases, respectively.

**D. The influence of local fields on the electrocaloric response of ferroelectric composites**

Noteworthy, the expression (8) for $\Delta T_{EC}$ can be used only if the electric field, penetrating in the BTO nanoparticles, is strong enough to change their spontaneous polarization significantly and to reach the coercive field, which is used in the estimates (8), and corresponding $\Delta T_{EC}$ is shown in **Figs. 2**. To study the role of local electric fields and to find the conditions when the external fields can effectively penetrate inside the BTO nanoparticles in the composite, we performed the FEM. To perform quick and large-scale 3D simulations for different polymer matrices, we assume that the relative dielectric permittivity of the nanoparticles changes continuously from the core value ($\varepsilon_c$) to the matrix value ($\varepsilon_M$). The change takes place in the diffuse shell which r-dependent permittivity, $\varepsilon_s(r)$, changes from $\varepsilon_c$ to $\varepsilon_M$ in accordance with a Gaussian law. The core permittivity includes the isotropic background permittivity ($\varepsilon_b$) and the anisotropic ferroelectric (or paraelectric) contributions, which are characterized by the effective value $\varepsilon_c = \varepsilon_b + \sqrt{\varepsilon_{11}^f \varepsilon_{33}^f}$. The assumption is oversimplified; however, it removes complications related to the mutual misorientation of the nanoparticle crystallographic axes in the composite and consideration of complicated and poorly known electric boundary conditions at the core-shell and shell-media interfaces (such as double electric layers). The estimate for $\varepsilon_{33}^f \approx 120$, $\varepsilon_{11}^f \approx 10^3$, and $\varepsilon_b = 4$ at room temperature gives the value $\varepsilon_c \approx 350$.

Two core-shell BTO nanoparticles, placed near the bottom electrode and surrounded by a dielectric matrix, are schematically shown in **Fig. 3(a)**. Distributions of external electric potential $\phi$, normal and lateral components of the quasi-static external electric field, $E_3^e$ and $E_1^e$, are shown in **Fig. 3(b)-(h)**. The voltage $V = 10$ V is applied to a 200-nm thick capacitor; $\varepsilon_c = 350$, $\varepsilon_M = 3$ (left column) and $\varepsilon_M = 30$ (right column). The diffuse shell is 5 nm thick, and the nanoparticle's radius is 20 nm. The remote top electrode is located at the 200 nm distance from the substrate. The component $E_3^e$ is much higher than $E_1^e$ inside the nanoparticles. Since $\varepsilon_c \gg \varepsilon_M$ the component $E_3^e$ in the nanoparticles is significantly smaller than the external field $E_3^0$ far from the nanoparticles, $E_3^0 = 50$ mV/nm. Namely, $E_3^e \approx 2$ mV/nm for $\varepsilon_M = 3$ and $E_3^e \approx 20$ mV/nm for $\varepsilon_M = 30$ inside the nanoparticles. However, the field $E_3^e \approx 20$ mV/nm is very close to the coercive field, $E_c \approx 22$ mV/nm, shown in **Fig. 1(b)-(d)**. Thus, we can conclude that the ratio $\varepsilon_M/\varepsilon_c$ should be higher than 0.1. Otherwise, the external field decreases so strongly inside the nanoparticle that it cannot significantly change the nanoparticle's spontaneous polarization.

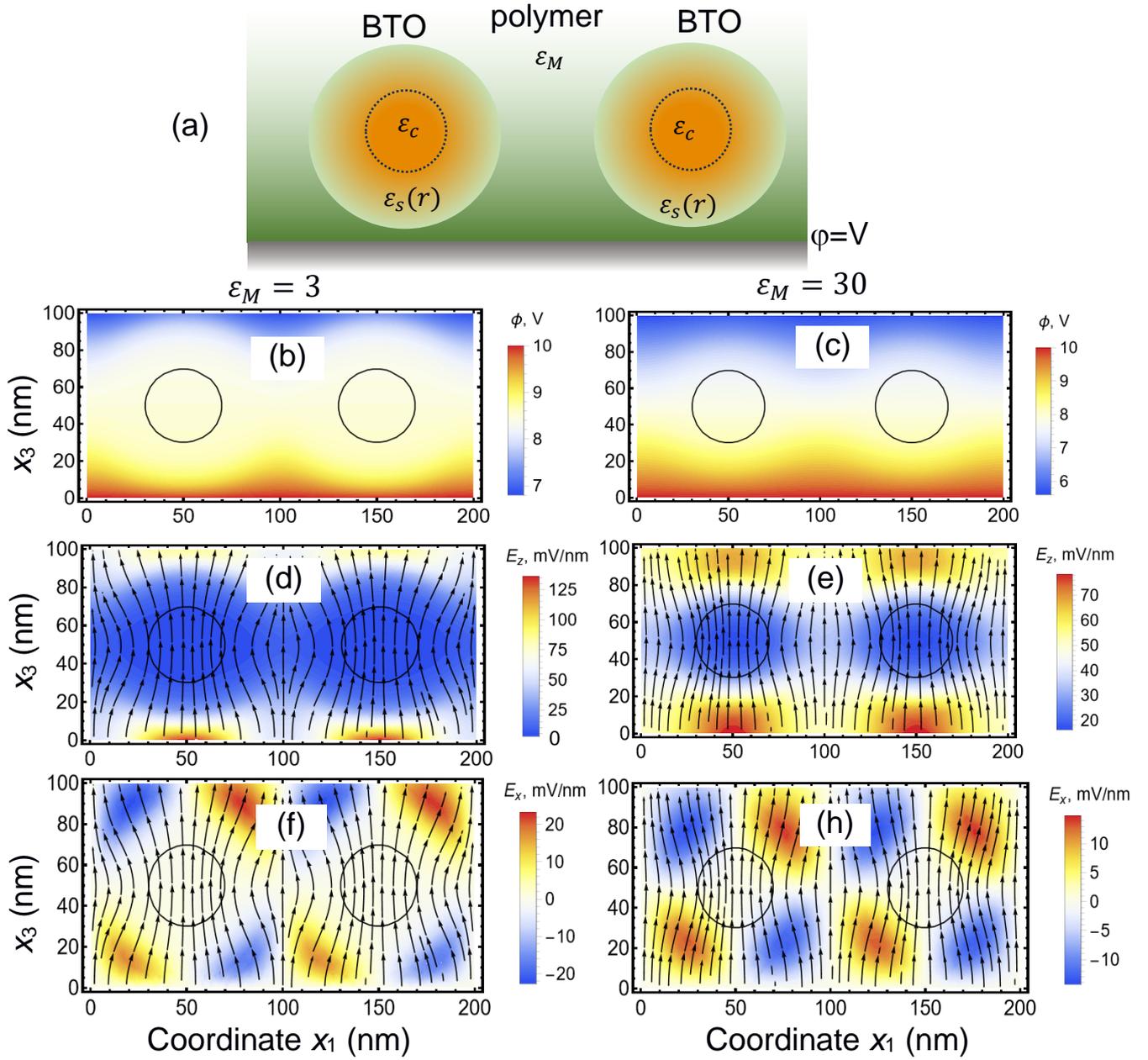

**FIGURE 3. (a)** Schematics of two BTO nanoparticles placed near the bottom electrode and surrounded by a dielectric matrix. Typical distributions of electric potential $\phi$ **(b, c)**, normal **(d, e)**, and lateral **(f, h)** components of the quasi-static external electric field in the system. Parameters: $V = 10$ V is applied to a 200-nm thick capacitor; $\varepsilon_c = 350$, $\varepsilon_M = 3$ (left column) and $\varepsilon_M = 30$ (right column). The radius of BTO nanoparticles is 20 nm, and their diffuse shell is 5 nm thick.

The enhancement of the local fields outside the nanoparticles exists near the biased electrode, where $\varphi \approx 10$ V and $E_3^e \approx 125$ mV/nm for $\varepsilon_M = 3$, or $E_3^e \approx 75$ mV/nm for $\varepsilon_M = 30$. Using the superposition principle of electric fields, we did not show the depolarization field induced by the spontaneous polarization in the nanoparticles in **Fig. 3**, because we aim to compare the external field components with the coercive field.

It is evident from the FEM results, which typical examples are shown in **Fig. 3**, that the effective dielectric permittivity of the composite ($\varepsilon_{eff}$) determines the degree of external field penetration inside the nanoparticles, and so the analytical estimates and self-consistent calculations of the $\varepsilon_{eff}$ dependence on the volume fraction of nanoparticles ($v_P$), dielectric constants of the matrix ($\varepsilon_M$) and the nanoparticles ($\varepsilon_P$) are in order. In **Appendix A3** we performed calculations according to the Landau approximation of a liner mixture [48], which is valid for $v_P \ll 0.1$, the Maxwell-Garnett model [49], which is suitable for $v_P$ less than 10%, the Bruggeman model [50], which is applicable $v_P <$(20-30)%, and the Lichtenecker-Rother model of a logarithmic mixture [51], which is valid for a random distribution of shapes and orientations of each component. Since any limitations for $v_P$ value are absent for the Lichtenecker-Rother model, the model seems the most suitable for further use because we consider dense composites with $v_P \geq 0.2$. According to the Lichtenecker-Rother model, the nanoparticle fraction $v_P$ must exceed 0.25 to increase the ratio $\varepsilon_{eff}/\varepsilon_c$ above 0.1 for the PVDF matrix, $v_P$ should be higher than 0.5 to reach $\varepsilon_{eff}/\varepsilon_c > 0.1$ in the PVB matrix, and $v_P > 0.55$ is required for $\varepsilon_{eff}/\varepsilon_c > 0.1$ for air or gaseous matrix (see **Fig. A1** in **Appendix A3** for details).

One also requires incorporating fillers with high heat conductance and relatively low heat capacitance (such as highly conducting or/and metallic nanoparticles), which act as heat sinks and facilitate the heat transfer from the ferroelectric nanoparticles to the polymer matrix. However, the presence of semiconducting or/and metallic nanoparticles leads to the appearance of the imaginary part of $\varepsilon_{eff}$, which is included neither in the effective media models nor in the FEM (see **Fig. 3**). Thus, it is principally important to determine the real and imaginary parts of $\varepsilon_{eff}$ correctly, and therefore, the next section is devoted to the experimental measurements of effective dielectric permittivity and screening characteristics for the dense composites consisting of quasi-spherical BTO nanoparticles incorporated in different organic polymers.

### III. EXPERIMENTAL DETERMINATION OF THE EFFECTIVE DIELECTRIC PERMITTIVITY AND LOSSES

#### A. Experimental results for the dense composite films and their discussion

Experimental studies have been carried out for the composite films consisting of $v_P =$(28 – 35) vol.% of the BTO nanoparticles incorporated in different organic polymers: polyvinyl-butyral (**PVB**) and ethyl-cellulose (**ETC**). The method of nanocomposites preparation by tape casting of suspensions based on BTO nanopowders is described in detail in Ref. [66].

The studied BTO nanoparticles were prepared by the non-isothermal solid-state synthesis. As it is seen from the TEM images (e.g., **Fig. 4(a)** and **Fig. B1** in **Appendix B**) the nanoparticles have a quasi-spherical shape with the diameter ranging from 17 nm to 47 nm, and the statistically averaged diameter being of 24 nm (see **Fig. 4(b)**). The XRD of as-prepared nanopowders reveals the content of the tetragonal barium titanate (space group P4mm) to be not less 99.9 wt. % (see **Fig. 4(c)**). The lattice constants determined by XRD are $a = b = 4.0126(4)$ Å and $c = 4.0428(8)$ Å, and the lattice tetragonality is $c/a = 1.007$. The average size of coherent scattering regions is 217(6) Å, which is in a reasonable agreement with the TEM results.

To create a suspension and layers of composites the nanosized and submicron BTO powders were used. The ethanol and isopropanol were used as media for wet grinding and disintegrants to prepare the polymer mixtures. The powdered polyvinyl butyral (PVB) Movital (Kuraray GmbH, Japan) brands B-98 and B75H with a molecular weight of 40.000 and 100.000, respectively, and ethyl-cellulose (ETC) (Merck GmbH, Germany) (viscosity of 5% solution in toluene/ethanol 80:20 − 100 cP) were used as binder polymers. The dibutyl phthalate (Merck GmbH, Germany) was used as a plasticizer.

The polymers PVB and ETC in the wet state as-prepared samples had the high-k dielectric permittivity, $\varepsilon_M \approx 16.9$ and 8.47, respectively [67, 68]. After short aging, the polymers got dry and had significantly lower static permittivities, $\varepsilon_M \approx 3.17$ and 3.35, respectively, in agreement with known data [69, 70]. The composite films were pressed on the cuprum-foiled surface of the fiberglass plate. The film thickness $h$ was 18 μm in the sample **TCS-27** with BTO nanoparticles having average size 24 nm in PVB and $h$=8.5 μm in the sample **TCS-34** with BTO submicron particles (average size 300 nm, Ferro corp., USA) in ETC. The silver electric contacts were made on the top film side either by thermal deposition or by paste application. The contacts have different surface areas. Their dimensions are 1x1, 1.5x1.5, 2x2, 2.5x2.5 and 3x3 mm. One of the samples with a set different contact areas made by thermal deposition is shown in **Fig.4(d)**. Each set with the same contact area consist of 3 to 7 samples. The electric transport in the studied samples occurs vertically through the composite film perpendicularly to its plane. The capacitance, loss-angle tangent, and active resistance in the parallel equivalent scheme were measured by LCR meters E7-12 and E12-1 at the frequencies 1 kHz, 100 kHz, and 1 MHz, and zero voltage bias. These measurements were performed at $T = 298$ K.

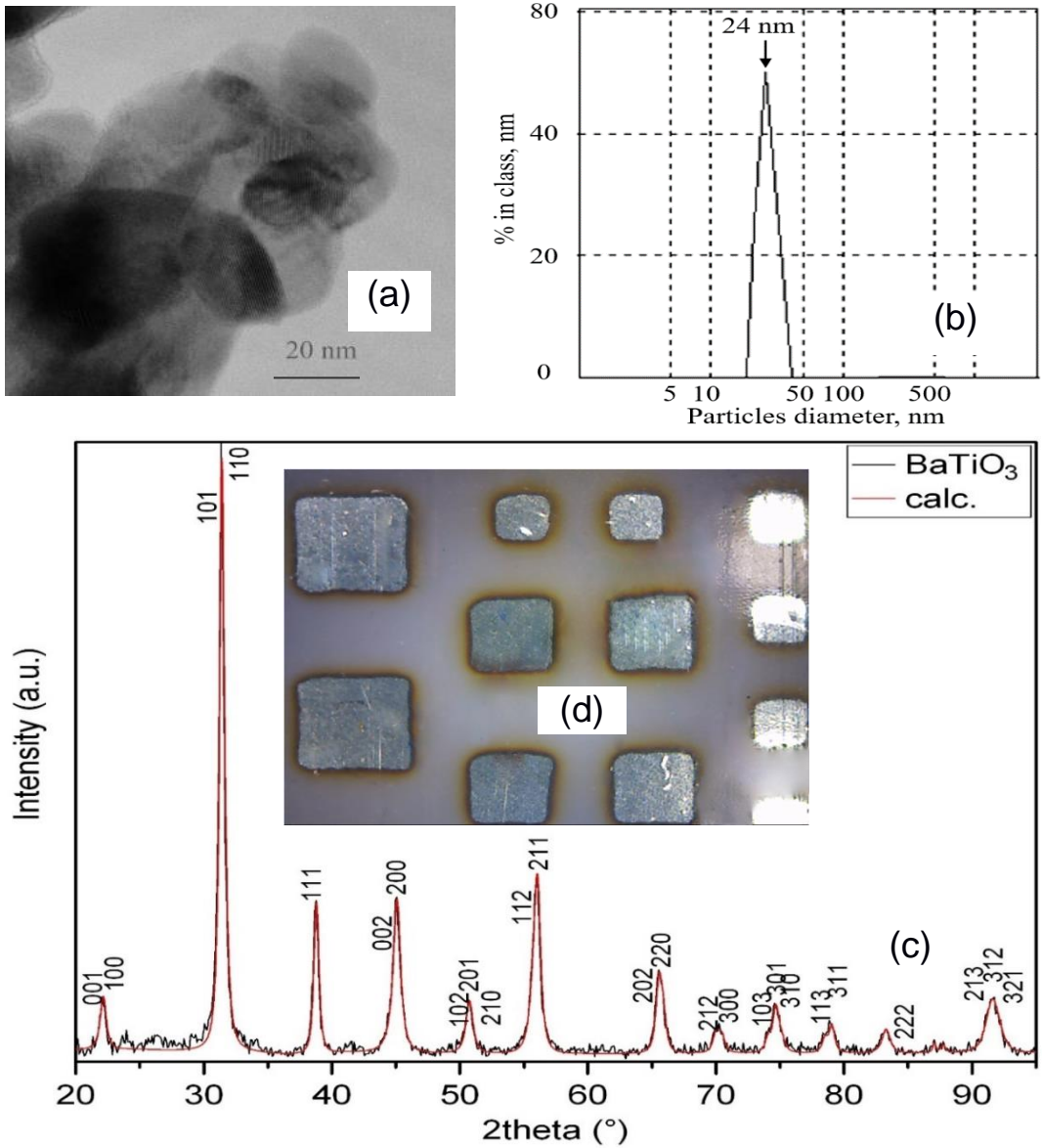

**FIGURE 4.** The TEM image of the BTO nanopowder **(a)**, granulometric composition of the BaTiO$_3$ powder (laser granulometry) **(b)**. XRD spectra of the as-prepared BaTiO$_3$ nanopowder, where the black curves are the spectra, and the red curve is fitting **(c)**, photo of a composite film on a foiled fiberglass with the top silver contacts **(d)**.

Also, we measured current-voltage (**I-V**) characteristics in the *dc* regime and the capacitance-voltage (**C-V**) characteristics in the dynamic *ac* mode. To avoid possible transient processes, the I-V measurements were done with a 1.5 s delay during and after each voltage step. The C-V measurements were made with voltage dynamically varying within the range ±10 V with the triangular and sinusoidal voltage scan. The dependences of the capacitance, loss-angle tangent, and active resistance vs top contact area are shown in **Fig. 5** and listed also numerically in **Table BII** in **Appendix B**. The I-V characteristics in the dc regime plotted in **Fig. B2** in **Appendix B**.

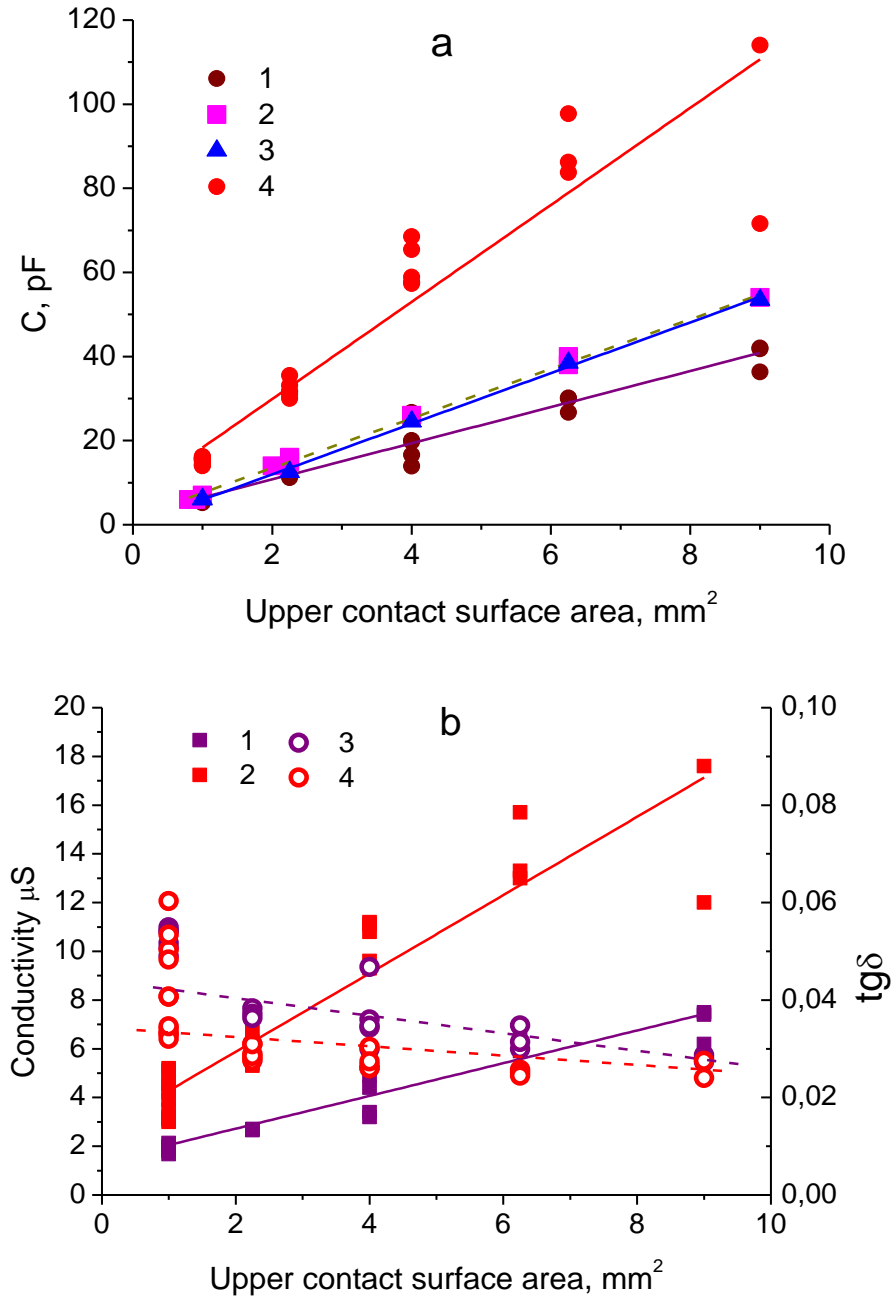

**FIGURE 5. (a)** Dependences of the composite film capacitance on the top contact area at 1 kHz, 600 kHz, and 1 MHz. The symbols 1, 2, and 3 correspond to the film TCS-27 (thickness $h = 18$ μm); the symbols 4 are for the film TCS-34 ($h = 8.5$ μm). Frequency: 1 and 4 - $f = 1$ MHz, 2 – $f = 1$ kHz, 3 – $f = 100$ kHz. The straight lines are the linear least squares fitting of experimental points. **(b)** Dependences of the conductivity (left scale, curves 1 and 2) and tangent of the losses angle (right scale, curves 3 and 4) vs. the top contact area. The symbols 1 and 3 correspond to the film TCS-27, 2 and 4 correspond to the film TCS-34.

Fitting the experimental data by the least squares method gives the specific capacitance $\langle C_f \rangle$ =5.89, 6.02 and 4.29 pF/mm² at 1 kHz, 100 kHz, and 1 MHz, respectively, for the composite TCS-27 and 11.54 pF/mm² at 1 MHz for the composite TCS-34. Using the expression $\varepsilon_{eff} = \frac{\langle C_f \rangle h}{\varepsilon_0 S}$, we calculated that the average relative dielectric permittivity of TCS-27 is equal to 11.98, 12.25, and 8.73 at 1 kHz, 100 kHz, and 1 MHz, respectively, and $\varepsilon_{eff} = 11.08$ for TCS-34 at 1 MHz. One should add that obtained samples have quite uniformly distributed inclusion by area. The most spread of the measured capacitance is observed at the largest area of the upper contact. At that, the largest mean squared error of the specific capacitance and other parameters determined by the least square method is less 7 %.

The typical I-V characteristics of the samples are shown in **Fig. B2**. The current density in fact linearly depends on the applied voltage, except for small loops in several cases, that evidences the possibility of samples to reveal hysteresis in I-V characteristics. The hysteresis behavior can be related to mixture of ferroelectricity and ionic response originating from diffusion of the silver paste into the sample. Therefore, it can be classified as the ferro-ionic response (see Ref. [71]). The studied composite samples in the steady-state regime of the vertical electric transport do not reveal any noticeable nonlinearities for the voltages less than 10 V. However, the experiment goal was the low-voltage linear regime to determine the effective media parameters, rather than the polarization switching.

To resume, we have determined the effective dielectric permittivity of the composite "BTO nanoparticles of (28 – 35) vol. % in the PVB or ETC matrices", which changes from 8.73 to 11.54 in the frequency range from $10^3$ to $10^6$ Hz. The conductivity of the composite film at that is low and keeps the tangent of the loss angle less than 0.05 in the films with different thicknesses at $f = 1$ MHz.

### B. Comparison of experimental results and theoretical models

Finally, let us compare the observed magnitude of the dielectric permittivity with the considered theoretical model. In accordance with our measurements $\varepsilon_{eff} = 8.73$ at $f = 1$ MHz, $\varepsilon_{eff} = 12.25$ at $f = 100$ kHz and $\varepsilon_{eff} = 11.98$ at $f = 1$ Hz for the composite TCS-27 at the room temperature, and $\varepsilon_{eff} = 11.08$ for the composite TCS-34 at 1 MHz. These values differ insignificantly because the frequency dispersion of BTO dielectric permittivity is negligibly small in the frequency range $f \leq 1$ MHz [72], and the value $\varepsilon_{eff} = 11.98$ measured at $f = 1$ Hz can be regarded as static. Using $v_P = 0.28$, $\varepsilon_M \approx 3.17$ (for PVB) or 3.35 (for ETC), and assuming $\varepsilon_P \approx \varepsilon_C \approx 350.4$, we estimate the value of effective permittivity within the effective medium models and obtained the following values of $\varepsilon_{eff}$:

$$\varepsilon_{eff}^L \approx 100, \quad \varepsilon_{eff}^{MG} \approx 7, \quad \varepsilon_{eff}^B \approx 14, \quad \varepsilon_{eff}^{LR} \approx 12. \qquad (11)$$

From Eq.(11), the value of $\varepsilon_{eff}^{L}$ estimated within the Landau model of the linear mixture overestimates the experimental result by an order of magnitude. The Maxwell-Garnett ($\varepsilon_{eff}^{MG}$) and Bruggeman ($\varepsilon_{eff}^{B}$) models underestimate (56%) or slightly overestimate (less than 8%), respectively, the experimental result. The value of $\varepsilon_{eff}^{LR} \approx 12$ estimated within the Lichtenecker-Rother model of logarithmic mixture very well agrees with the experimental result $\varepsilon_{eff} = 11.98$ measured at $f = 1$ Hz. Note, that the Lichtenecker-Rother model allows us to consider the presence of the third fraction (e.g., fillers) in the composite by adding the third component to the mixture (see Eq.(A.12)).

## IV. THE ELECTROCALORIC RESPONSE OF FERROELECTRIC COMPOSITES

### A. Analytical estimates of the EC cooling

Since the application of the electric field during the EC heating and cooling are adiabatic processes, the EC response of a ferroelectric composite consisting of the BTO nanoparticles incorporated in a heat conductive media can be estimated from the heat-balance equation, listed in **Appendix C**:

$$\Delta T_{P-M} = \frac{v_P \rho_P C_P}{v_P \rho_P C_P + v_M \rho_M C_M} \Delta T_{EC}. \tag{12}$$

Here $\Delta T_{P-M}$ is the temperature change of the composite induced by the EC effect in the BTO nanoparticles, $v_P$ and $v_M = 1 - v_P$ are the relative volumes of the nanoparticles and the media, respectively, $\rho_P$ and $\rho_M$ are their volume densities, $C_P$ and $C_M$ are their specific heat capacitances, and $T$ is the initial temperature of the media.

As one can see from Eq.(12), $|\Delta T_{P-M}| \leq |\Delta T_{EC}|$, and thus, it makes sense to prepare very dense composites with $v_P \geq v_M$ and/or to use polymers and/or fillers with lower heat mass $\rho_M C_M \leq \rho_P C_P$. Below, we analyze the numerical estimates of $\Delta T_{P-M}$ for dense composites, whose experimental results were presented in the previous section. The material parameters of the composites are listed in **Appendix C.**

The dependence of $\Delta T_{P-M}$ on the BTO particle fraction $v_P$ and their EC cooling temperature $\Delta T_{EC}$, shown in **Fig. 6**, is almost the same for all three polymers, PVB, PVDF and ETC, because $\rho_P C_P \gg \rho_M C_M$ for the polymers as. To achieve a significant EC cooling, $\Delta T_{P-M} < -5$ K, the fraction $v_P$ should be higher than 0.25 for the average core radius 6 nm, and higher than 0.85 for the average core radius 60 nm. For the BTO nanopowders in the air, the product $\rho_{Air} C_{Air}$ is about $10^3$ times smaller than the product $\rho_P C_P$, Therefore, the dependence of $\Delta T_{P-M}$ on $v_P$ and $\Delta T_{EC}$, calculated for the BTO nanopowders and shown in **Fig. C1**, is almost $v_P$-independent and looks very different from **Fig. 6(a)** and **6(b)**, because $\Delta T_{P-M} \approx \Delta T_{EC}$ for $0.001 < v_P \leq 1$.

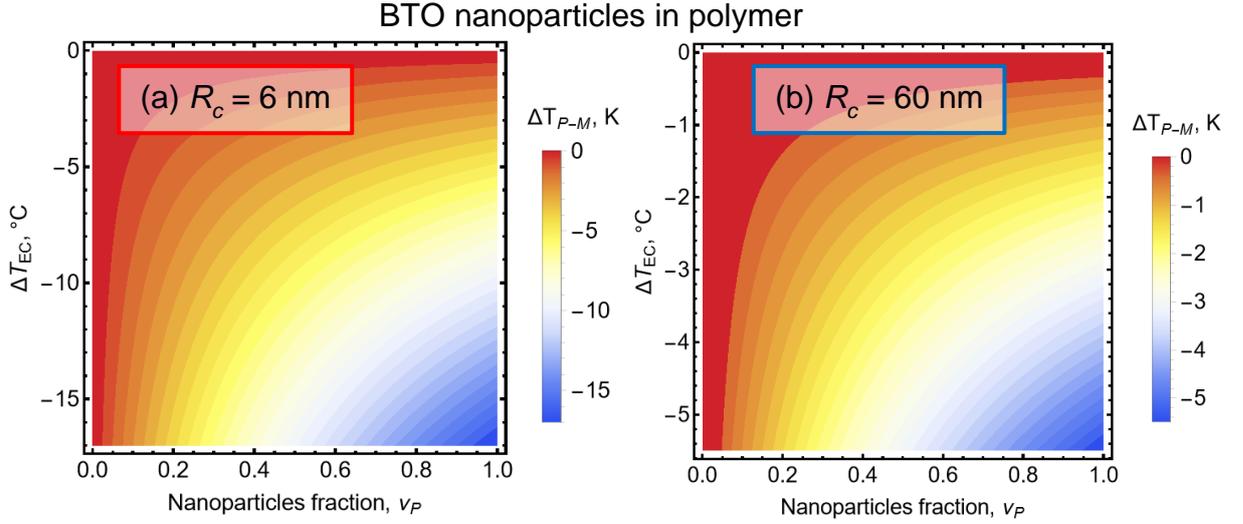

**FIGURE 6.** The dependence of $\Delta T_{P-M}$ on the nanoparticle fraction $v_P$ and EC cooling temperature $\Delta T_{EC}$ calculated for BTO nanoparticles with the core radius $R_c = 6$ nm **(a)** and 60 nm **(b)** embedded in different polymers, such as PVB, PVDF, or ETC.

Let us underline that the simple expression (12) has a practical sense only if the heat flux from ferroelectric nanoparticles can be transferred enough rapidly to the effective media and then from the media to the working substance, which should be cooled or heated by the EC effect in the particles. Thus, neither air, nor any other media with low heat conductance is not suitable for use as the composite. It is reasonable to incorporate fillers with high heat conductance and relatively low heat capacitance (such as metallic nanoparticles), which act as heat sinks and can facilitate the heat transfer from the ferroelectric nanoparticles to the effective media and from the media to the working substance.

**B. Advanced applications of dense ferroelectric nanocomposites for EC cooling**

There is a lot of useful applications of electrocaloric materials for cooling, including cooling at low working temperatures. A perspective goal is to combine several useful functions of the EC material for its advanced applications in modern devices. It is required to avoid any moving parts for durability, to reach tunability of working temperatures, and to use easy applicable electric fields for cooling instead of large magnetic fields for magnetocaloric cooling, which application is often problematic and hardly scalable.

The EC ferroelectric composite can be polarized and repolarized, so it, like any ferroelectric, is suitable for use as non-volatile memory cells and/or other memory elements. For example, a thin ferroelectric layer can serve as a dielectric gate oxide of a field-effect transistor with a graphene channel [73, 74], where the time-dependent transport characteristics of graphene are tuned by the ferroelectric polarization and interface charge trapping [ 75 ]. Also, recent advances in miniaturization of ferroelectric EC elements can open their use for cooling of various electronic elements, including critical electronics and processors (instead of fans and Peltier elements).

According to requirements for the state-of-the-art of the 3-nm technology of integrated circuits, the modern design of the transistor suggests that the "Gate-All-Around" (**GAA**) should control the stack of channels. The cooling of the stack becomes urgent because the total heating rapidly increases with the number of channels in the stack. Thus, an interesting combination can be a solid-state high-temperature superconductor (**HTSC**), acting as a channel (or stack of channels) of a field effect transistor (**FET**) [76], and a dense ferroelectric nanocomposite with a high EC effect, acting as the gate dielectric controlling the channel(s) conductivity [see **Fig. 7(a)**]. The good tunability of EC cooling temperature due to the strain and size effects in ferroelectric nanoparticles allows to select the operating temperature $T$ in the wide range (50 – 300) K (see green and blue areas in **Fig. 2(j)-(l)** for the EC response). If $T$ is slightly above the temperature of the superconductive phase transition $T_{SC}$ in the HTSC, the channel conductivity can be switched to the superconducting state by the application of electric field to the EC composite. The nanocomposite can be cooled under the adiabatic application of an electric voltage below (or close) the coercive value followed by a slow non-adiabatic (e.g., isothermal) "reset" of the voltage (e.g., to zero value). The sign of the voltage does not matter since the EC effect is proportional to even degrees of polarization in accordance with Eq.(10) [see **Figs. 7(b)** and **7(c)**]. However, it does matters whether the nanocomposite manages to cool the HTSC at least on the (5 – 10) degrees, until its spontaneous polarization is restored back during the field reset. The voltage application protocol with sharp rises and smooth relaxation is not a problem, but the limitations on the working frequency can be more severe. A metallic cover, which is also highly heat-conductive, can protect the HTSC channels from the electric field penetrating through the nanocomposite and would have minimal influence on the heat transfer.

Anyway, the modulation of superconducting current can be used is some types of logic keys, triggers and higher harmonic generators [76] operating at relatively low frequencies, required for successful EC cooling of the channel by the ferroelectric nanocomposite. The operating temperature $T$ also matters. If the temperature of HTSC superconductive transition is close to the liquid nitrogen temperature or higher, it becomes possible to use the considered BTO nanoparticles in polymer matrices, because the predicted maxima of the giant EC cooling effect in the single nanoparticle corresponds to relatively low temperatures and high tensile Vegard stresses (see green and blue areas in **Fig. 2(j)-(l)** for the EC response, which correspond to the wide temperature range 50 – 300 K).

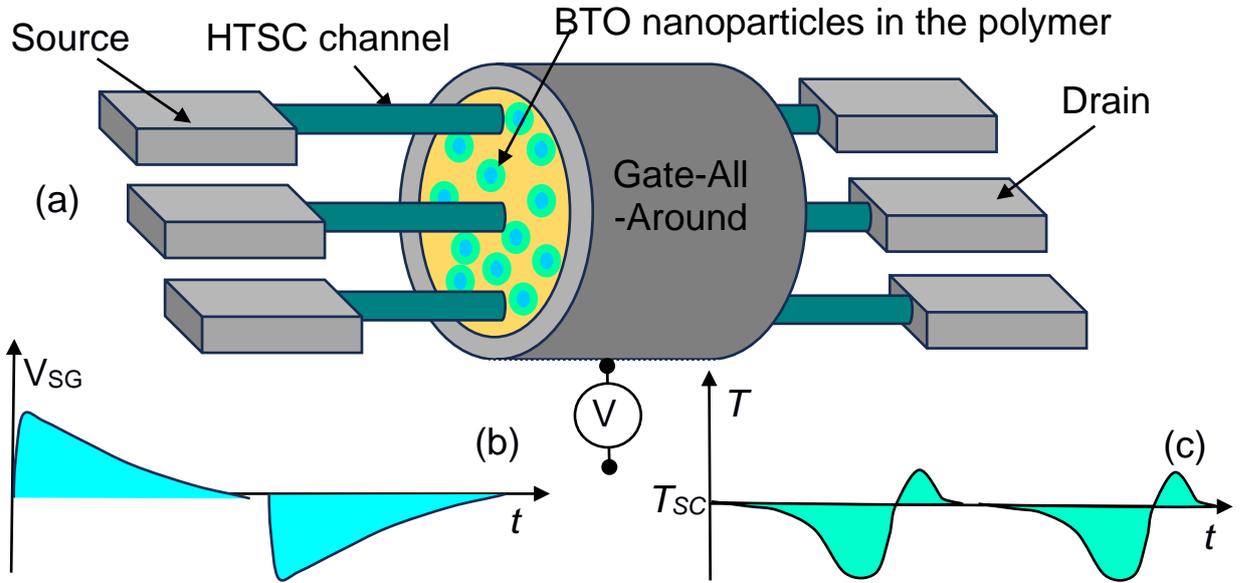

**FIGURE 7. (a)** A scheme of the FET with a stack of HTSC channels covered by the GAA dielectric made of the ferroelectric nanocomposite with a high EC effect. The application protocol of the gate voltage $V_{SG}(t)$ **(b)** and the EC cooling of the nanocomposite induced by the voltage change **(c)**.

Notably, that the re-polarization of the whole nanocomposite is not required for the FET operation. It rather should be avoided, because a strong depolarization electric field could influence on the HTSC conductivity, and thus we propose to screen the channels. Instead, it is necessary to modulate the polarization of individual nanoparticles and to reach their maximal EC cooling. Thus, the dense ferroelectric nanocomposites, we prepared and studied experimentally, are suitable for the purpose.

The advantages of the studied nanocomposites are the good tunability of EC cooling temperature due to the stresses and size effects in the BTO nanoparticles, high radiation and temperature stability, the absence of moving parts, and the easy electric field control of the EC cooling. The disadvantage is a relative low EC cooling effect in a bulk BTO (~2.5 K), which is more than in 10 times smaller than maximal cooling temperatures (~25 – 50 K) achieved in e.g., lead-containing ferroelectric relaxors and their thin layers of thickness (200 – 300) nm [77, 78]. The disadvantage can be compensated by using ultra-small (size ~ 10 nm) stressed BTO nanoparticles, which EC cooling can reach (15 – 20) K. They are lead-free, easy for preparation and well-suitable for integration into flexible polymers with minimal thickness (100 – 200) nm.

## V. CONCLUSIONS

Using the analytical LGD approach, we calculated the spontaneous polarization $P_s$, PE coefficient $\Pi_s$, relative dielectric permittivity $\varepsilon_f$, and the EC temperature change $\Delta T_{EC}$ of a single-domain BTO nanoparticle placed in a polymer matrix. The elastic defects concentrated near the surface induce strong Vegard stresses can reach several GPa and strongly stress the nanoparticle.

Due to the size effects and Vegard stresses, $P_s$ can reach 50 μC/cm$^2$ for stretched BTO nanoparticles with sizes (10 – 25) nm, and their EC cooling temperature $\Delta T_{EC}$ is more than 17 K far from the boundary of the size-induced paraelectric-ferroelectric phase transition. Therefore, the synergy of size effects and Vegard stresses can significantly improve the spontaneous polarization (1.5 – 2 times) and enhance the EC properties (2 – 7 times) of the BTO nanoparticles in comparison with a bulk BTO and unstrained nanoparticles.

Experimental studies have been performed for the dense nanocomposites consisting of (28 – 35) vol.% BTO nanoparticles incorporated in the PVB and ETC polymers and covered by Ag electrodes. We measured the current-voltage, capacitance-voltage and charge-voltage characteristics of the nanocomposites to compare with the effective media models. To reach the maximal EC response, it makes sense to prepare dense composites with a nanoparticles volume ratio about or more 30% and to use polymers with lower heat mass than the nanoparticles.

Due to the size effects, BTO nanoparticles spontaneously stressed by elastic defects, such as oxygen vacancies or other elastic dipoles, can cause the high EC response of the dense ferroelectric nanocomposites. The conclusion is general for other ferroelectric nanoparticles spontaneously stressed by inherent (or incorporated) elastic defects, which create enough strong chemical pressure and so cause their giant EC response.

The advantages of the studied lead-free ferroelectric nanocomposites are the good tunability and enhancement of EC cooling temperature due to the size effects and Vegard stresses in the ferroelectric nanoparticles, relatively high EC cooling temperature and easy electric field control. These advantages make promising the nanocomposites for cooling of conventional and innovative electronic elements, such as FETs with high-temperature superconductor channels.

**Acknowledgements.** The theoretical part of the research, materials preparation characterization is sponsored by the NATO Science for Peace and Security Programme under grant SPS G5980 "FRAPCOM" (A.N.M, L.P.Y., S.I. and Z.K.). Measurements are sponsored in part by the Target Program of the National Academy of Sciences of Ukraine, Project No. 4.8/23-p "Innovative materials and systems with magnetic and/or electrodipole ordering for the needs of using spintronics and nanoelectronics in strategically important issues of new technology" (O.S.P., N.V.M., V.N.P., and V.V.V.).

**Authors contribution.** A.N.M. generated the research idea, formulated the problem, performed analytical calculations, and wrote the paper draft. E.A.E., H.V.S., L.K. and L.P.Y. wrote the codes and prepared figures. S.I. and O.V.S prepared the nanoparticles and nanocomposites and characterized them. O.S.P., N.V.M., V.N.P., and V.V.V. performed experiments, and V.V.V. wrote the experimental part. A.N.M., Z.K. and V.V.V. worked on the results interpretation, discussion, and paper improvement.

**APPENDIX A. Mathematical formulation of the problem and computation details**

**A1. Electric field, polarization, and elastic field in the ferroelectric nanoparticle**

The nanoparticle geometry is shown in **Fig. 1(a)** of the main text. The electric displacement vector has the form $\boldsymbol{D} = \varepsilon_0 \boldsymbol{E} + \boldsymbol{P}$ inside the nanoparticle. In this expression, $\boldsymbol{P}$ is an electric polarization containing the spontaneous and field-induced contributions. The expression $D_i = \varepsilon_0 \varepsilon_e E_i$ is valid in the isotropic effective medium.

The electric field components, $E_i$, are derived from the electric potential $\varphi$ in a conventional way, $E_i = -\partial\varphi/\partial x_i$. The potential $\varphi_f$ satisfies the Poisson equation in the ferroelectric nanoparticle (subscript "*f*"):

$$\varepsilon_0 \varepsilon_b \left( \frac{\partial^2}{\partial x_1^2} + \frac{\partial^2}{\partial x_2^2} + \frac{\partial^2}{\partial x_3^2} \right) \varphi_f = \frac{\partial P_i}{\partial x_i} - \rho(\vec{r}, \varphi_f), \qquad 0 \leq r \leq R_s. \qquad (A.1a)$$

Here $\varepsilon_0$ is the universal dielectric constant, $\varepsilon_b$ is a relative background permittivity of the core [79]. As a rule, $4 < \varepsilon_b < 10$. $\rho(\vec{x}, \varphi_f)$ is the free charge density, and $r = \sqrt{x_1^2 + x_2^2 + x_3^2}$. For FEM, we assume that the Debye-Huckel approximation is valid for the charge density in the screening shell, $\rho(\vec{r}, \varphi_f) = n_0(\vec{r}) \sinh\left(\frac{e\varphi_f}{k_B T}\right) \approx -\varepsilon_0 \frac{\varphi_f}{\lambda_0(\vec{r})^2}$, where $\lambda_0(\vec{r}) = \sqrt{\frac{\varepsilon_0 k_B T}{e^2 n_0(\vec{r})}}$ is the "net" Debye-Huckel screening length, which value is defined by the concentration of free carriers $n_0(\vec{r})$ in the diffuse shell. As a rule, one can use any well-localized function for $n_0(\vec{r})$ as interpolation, e.g., $n_0(\vec{r}) = n_0 \exp\left[-\frac{(r-R_m)^2}{\Delta R^2}\right]$, where $R_m = \frac{R_s - R_c}{2}$ and $\Delta R \ll R_c$. The dependence of the core polarization on the electric field can be represented as $\vec{P} = \{\varepsilon_0 \chi_{11} E_1, \ \varepsilon_0 \chi_{22} E_2, \ \varepsilon_0 \chi_{33} E_3 + P_s(\vec{r})\}$, where $\chi_{ii}$ is the linear dielectric susceptibility, $P_s(\vec{r})$ is the core spontaneous polarization directed along the polar axis z. Since the linear dielectric permittivity $\varepsilon_{ii} = \chi_{ii} + \varepsilon_b$, Eq.(A.1a) can be approximated as:

$$\left( \varepsilon_{11} \frac{\partial^2}{\partial x_1^2} + \varepsilon_{22} \frac{\partial^2}{\partial x_2^2} + \varepsilon_{33} \frac{\partial^2}{\partial x_3^2} \right) \varphi_f \approx \frac{1}{\varepsilon_0} \frac{\partial P_s}{\partial x_3} - \frac{\varphi_f}{\lambda_0(r)^2}, \qquad 0 < r \leq R_s, \qquad (A.1b)$$

Next, Eq.(A.1b) can be rewritten as

$$\left( \frac{\partial^2}{\partial x_1^2} + \frac{\partial^2}{\partial x_2^2} + f_a \frac{\partial^2}{\partial x_3^2} \right) \varphi_f \approx \frac{1}{\varepsilon_0 \varepsilon_{11}} \frac{\partial P_s}{\partial x_3} - \frac{\varphi_f}{\lambda(r)^2}, \qquad 0 < r \leq R_s, \qquad (A.1c)$$

where $f_a = \frac{\varepsilon_{33}}{\varepsilon_{11}}$ is the factor of dielectric anisotropy and $\lambda(\vec{r}) = \sqrt{\frac{\varepsilon_0 \varepsilon_{11} k_B T}{e^2 n_0(\vec{r})}}$ is the "dressed" Debye-Huckel screening length, which value is defined by the concentration of free carriers $n_0(\vec{r})$ in the diffuse shell and its paraelectric permittivity $\varepsilon_{11} = \varepsilon_{22}$.

The electric potential $\varphi_e$ in the external media outside the particle satisfies the Laplace equation (subscript "*e*"):

$$\varepsilon_0 \varepsilon_M \left( \frac{\partial^2}{\partial x_1^2} + \frac{\partial^2}{\partial x_2^2} + \frac{\partial^2}{\partial x_3^2} \right) \varphi_e = 0, \qquad r > R_s. \qquad (A.1d)$$

Equations (A.1) are supplemented with the continuity conditions for electric potential and normal components of the electric displacements at the particle surface:

$$(\varphi_e - \varphi_f)|_{r=R_s} = 0, \quad \vec{n}(\vec{D}_e - \vec{D}_f)|_{r=R_s} = 0, \tag{A.1e}$$

Either charges are absent, or the applied voltage is fixed at the boundaries of the computation region:

$$\frac{\partial \varphi_e}{\partial x_l} n_l \bigg|_{x=\pm\frac{L}{2}} = 0, \quad \frac{\partial \varphi_e}{\partial x_l} n_l \bigg|_{y=\pm\frac{L}{2}} = 0, \quad \varphi_e|_{z=+\frac{L}{2}} = 0, \quad \varphi_e|_{z=-\frac{L}{2}} = V_e. \tag{A.1f}$$

Here $V_e$ is the applied voltage difference, and $L$ is the size of the computation region.

The LGD free energy functional $G$ of the core polarization $\mathbf{P}$ additively includes a Landau expansion on the 2-nd, 4-th, 6-th and 8-th powers of the polarization, $G_{Landau}$; a polarization gradient energy contribution, $G_{grad}$; an electrostatic contribution, $G_{el}$; the elastic, linear, and nonlinear electrostriction couplings and flexoelectric contributions, $G_{es+flexo}$; and a surface energy, $G_S$. The functional $G$ has the form [80, 81, 82]:

$$G = G_{Landau} + G_{grad} + G_{el} + G_{es+flexo} + G_{VS} + G_S, \tag{A.2a}$$

$$G_{Landau} = \int_{0<r<R_c} d^3r \left[ a_i P_i^2 + a_{ij} P_i^2 P_j^2 + a_{ijk} P_i^2 P_j^2 P_k^2 + a_{ijkl} P_i^2 P_j^2 P_k^2 P_l^2 \right], \tag{A.2b}$$

$$G_{grad} = \int_{0<r<R} d^3r \frac{g_{ijkl}}{2} \frac{\partial P_i}{\partial x_j} \frac{\partial P_k}{\partial x_l}, \tag{A.2c}$$

$$G_{el} = -\int_{0<r<R_c} d^3r \left( P_i E_i + \frac{\varepsilon_0 \varepsilon_b}{2} E_i E_i \right) - \frac{\varepsilon_0}{2} \int_{R_c<r<R_s} \varepsilon_{ij}^s E_i E_j d^3r - \frac{\varepsilon_0}{2} \int_{r>R+\Delta R} \varepsilon_{ij}^e E_i E_j d^3r, \tag{A.2d}$$

$$G_{es+flexo} = -\int_{0<r<R_c} d^3r \left( \frac{s_{ijkl}}{2} \sigma_{ij} \sigma_{kl} + Q_{ijkl} \sigma_{ij} P_k P_l + Z_{ijklmn} \sigma_{ij} P_k P_l P_m P_n + \frac{1}{2} W_{ijklmn} \sigma_{ij} \sigma_{kl} P_m P_n + F_{ijkl} \sigma_{ij} \frac{\partial P_l}{\partial x_k} \right), \tag{A.2e}$$

$$G_S = \frac{1}{2} \int_{r=R_c} d^2r \, a_{ij}^{(S)} P_i P_j. \tag{A.2f}$$

The coefficient $a_i$ linearly depends on temperature $T$:

$$a_i(T) = \alpha_T [T - T_C(R_c)], \tag{A.3a}$$

where $\alpha_T$ is the inverse Curie-Weiss constant, and $T_C(R_c)$ is the ferroelectric Curie temperature renormalized by electrostriction and surface tension as [80, 81]:

$$T_C(R_c) = T_C \left( 1 - \frac{Q_c}{\alpha_T T_C} \frac{2\mu}{R_c} \right), \tag{A.3b}$$

where $T_C$ is a Curie temperature of a bulk ferroelectric. $Q_c$ is the sum of the electrostriction tensor diagonal components, which is positive for most ferroelectric perovskites with cubic m3m symmetry in the paraelectric phase, namely $0.005 < Q_c < 0.05$ (in m$^4$/C$^2$). $\mu$ is the surface tension coefficient.

Tensor components $a_{ij}$, $a_{ijk}$ and $a_{ijkl}$ are listed in **Table AI.** The gradient coefficients tensor $g_{ijkl}$ are positively defined and regarded as temperature-independent. The following designations are used in Eq.(A.2e): $\sigma_{ij}$ is the stress tensor, $s_{ijkl}$ is the elastic compliances tensor, $Q_{ijkl}$, $Z_{ijklmn}$ and $W_{ijklmn}$ are the linear and two nonlinear electrostriction tensors, whose values and/or ranges are listed in **Table AI**.

Since μ is relatively small, not more than (1 – 4) N/m for most perovskites, and to focus on the influence of linear and nonlinear electrostriction effects, we do not consider the surface tension and flexoelectric coupling in this work and put $\mu = 0$ and $F_{ijkl} = 0$.

Allowing for the Khalatnikov mechanism of polarization relaxation [83], minimization of the free energy (A.2) with respect to polarization leads to three coupled time-dependent Euler-Lagrange equations for polarization components inside the core,

$$\frac{\delta G}{\delta P_i} = -\Gamma \frac{\partial P_i}{\partial t}, \tag{A.4}$$

Where the subscript $i = 1, 2, 3$, $\Gamma$ and is the temperature-dependent Khalatnikov coefficient [84].

The boundary condition for polarization at the core-shell interface $r = R_c$ is:

$$a_{ij}^{(S)} P_j + g_{ijkl} \frac{\partial P_k}{\partial x_l} n_j \Big|_{r=R_c} = 0, \tag{A.5}$$

where $n_j$ are the components of the outer normal to the surface, and the subscripts $\{i, j, k, l\} = \{1, 2, 3\}$. Below, we use the so-called "natural" boundary conditions corresponding to $a_{ij}^{(S)} = 0$, which support the formation of a single-domain state in the core.

Elastic stresses satisfy the equation of mechanical equilibrium in the computation region,

$$\frac{\partial \sigma_{ij}}{\partial x_j} = 0, \qquad -L/2 < \{x, y, z\} < L/2. \tag{A.6}$$

Elastic equations of state follow from the variation of the energy (A.2e) with respect to elastic stress, $\frac{\delta G}{\delta \sigma_{ij}} = -u_{ij}$, namely:

$$s_{ijkl}\sigma_{kl} + Q_{ijkl}P_k P_l + Z_{ijkl}P_k P_l P_m P_n + W_{ijklmn}\sigma_{kl}P_m P_n = u_{ij}, \tag{A.7}$$

where $0 < r \leq R_c$, $u_{ij}$ is the strain tensor components related to the displacement components $U_i$ in the following way: $u_{ij} = (\partial U_i/\partial x_j + \partial U_j/\partial x_i)/2$.

The elastic displacement components $U_i$ and normal stresses $\sigma_{ij}$ are continuous functions at the core-shell interface ($r = R_c$):

$$U_i|_{r=R_c-0} = U_i|_{r=R_c+0}, \qquad \sigma_{ij}n_j\big|_{r=R_c-0} = \sigma_{ik}n_k|_{r=R_c+0}, \tag{A.8a}$$

as well as at the interface between the shell and the external media ($r = R_s$):

$$U_i|_{r=R_s-0} = U_i|_{r=R_s+0}, \qquad \sigma_{ij}n_j\big|_{r=R_s-0} = \sigma_{ik}n_k|_{r=R_s+0}. \tag{A.8b}$$

**Table AI.** LGD coefficients and other material parameters of a BaTiO$_3$ core in Voigt notations. Adapted from Ref. [32].

| Parameter, its description, and dimension (in the brackets) | The numerical value or variation range of the LGD parameters |
|---|---|
| Expansion coefficients $a_i$ in the term $a_i P_i^2$ in Eq.(A.2b) (C$^{-2}\cdot$mJ) | $a_1 = 3.33(T-383)\times 10^5$ |
| Expansion coefficients $a_{ij}$ in the term $a_{ij}P_i^2 P_j^2$ in Eq.(A.2b) (C$^{-4}\cdot$m$^5$J) | $a_{11} = 3.6\,(T-448)\times 10^6$, $a_{12} = 4.9\times 10^8$ |
| Expansion coefficients $a_{ijk}$ in the term $a_{ijk}P_i^2 P_j^2 P_k^2$ in Eq.(A.2b) (C$^{-6}\cdot$m$^9$J) | $a_{111} = 6.6\times 10^9$, $a_{112} = 2.9\times 10^9$, $a_{123} = 3.64\times 10^{10}+7.6(T-293)\times 10^{10}$. |
| Expansion coefficients $a_{ijkl}$ in the term $a_{ijkl}P_i^2 P_j^2 P_k^2 P_l^2$ in Eq.(A.2b) (C$^{-8}\cdot$m$^{13}$J) | $a_{1111} = 4.84\times 10^7$, $a_{1112} = 2.53\times 10^7$, $a_{1122} = 2.80\times 10^7$, $a_{123} = 9.35\times 10^7$. |
| Linear electrostriction tensor $Q_{ijkl}$ in the term $Q_{ijkl}\sigma_{ij}P_k P_l$ in Eq.(A.2e) (C$^{-2}\cdot$m$^4$) | In Voigt notations $Q_{ijkl} \rightarrow Q_{ij}$, which are equal to $Q_{11}=0.11$, $Q_{12}=-0.045$, $Q_{44}=0.059$. For a spherical nanoparticle, the combinations and functions of $Q_{ij}$ are used: $Q_c = \frac{Q_{11}+2Q_{12}}{3} = 0.0067$. |
| Nonlinear electrostriction tensor $Z_{ijklmn}$ in the term $Z_{ijklmn}\sigma_{ij}P_k P_l P_m P_n$ in Eq.(A.2e) (C$^{-4}\cdot$m$^8$) | In Voigt notations $Z_{ijklmn} \rightarrow Z_{ijk}$. For a spherical nanoparticle, the combinations and functions of $Z_{ijk}$ are used: $Z_c = Z_{111} + 2Z_{211}$, $Z_c$ varies in the range $-1 \leq Z_c \leq 1$ as free parameter |
| Nonlinear electrostriction tensor $W_{ijklmn}$ in the term $W_{ijklmn}\,\sigma_{ij}\sigma_{kl}P_m P_n$ in Eq.(A.2e) (C$^{-2}\cdot$m$^4$ Pa$^{-1}$) | In Voigt notations $W_{ijklmn} \rightarrow W_{ijk}$. For a spherical nanoparticle, the combinations and functions of $W_{ijk}$ are used: $W_c = W_{111} + 2W_{112} + 2W_{123} + 4W_{122}$, $W_c$ varies in the range of $0 \leq W_c \leq 10^{-12}$ as free parameter |
| Elastic compliances tensor, $s_{ijkl}$, in Eq.(A.2e) (Pa$^{-1}$) | In Voigt notations $s_{ijkl} \rightarrow s_{ij}$, which are equal to $s_{11}=8.3\times 10^{-12}$, $s_{12}=-2.7\times 10^{-12}$, $s_{44}=9.24\times 10^{-12}$. |
| Polarization gradient coefficients $g_{ijkl}$ in Eq.(A.2c) (C$^{-2}$m$^3$J) | In Voigt notations $g_{ijkl} \rightarrow g_{ij}$, which are equal to: $g_{11}=1.0\times 10^{-10}$, $g_{12}= 0.3\times 10^{-10}$, $g_{44}= 0.2\times 10^{-10}$. |
| Surface energy coefficients $a_{ij}^{(S)}$ in Eq.(A.2f) | 0 (that corresponds to the natural boundary conditions) |
| Core radius $R_c$ (nm) | Variable: 5 – 50 |
| Background permittivity $\varepsilon_b$ in Eq.(A.2d) (unity) | 7 |

\* $\alpha = 2a_1$, $\beta = 4a_{11}$, $\gamma = 6a_{111}$, and $\delta = 8a_{1111}$

## A2. The dynamics of quasi-static polarization, PE coefficient, dielectric susceptibility and EC response in the external field

The dynamics of quasi-static polarization $P_3$, PE coefficient $\Pi_3$ and dielectric susceptibility $\chi_{33}$ in the external field $E_3$ follows from the time-dependent LGD equations, which have the form:

$$\Gamma\frac{\partial P_3}{\partial t} + [\alpha - 2\sigma_i(Q_{i3} + W_{ij3}\sigma_j)]P_3 + (\beta - 4Z_{i33}\sigma_i)P_3^3 + \gamma P_3^5 + \delta P_3^7 - g_{33kl}\frac{\partial^2 P_3}{\partial x_k \partial x_l} = E_3,$$
(A.9a)

$$\Gamma\frac{\partial \chi_{33}}{\partial t} + [\alpha - 2\sigma_i(Q_{i3} + W_{ij3}\sigma_j) + 3(\beta - 4Z_{i33}\sigma_i)P_3^2 + 5\gamma P_3^4 + 7\delta P_3^6]\chi_{33} - g_{33kl}\frac{\partial^2 \chi_{33}}{\partial x_k \partial x_l} = 1,$$
(A.9b)

$$\Gamma\frac{\partial \Pi_3}{\partial t} + [\alpha - 2\sigma_i(Q_{i3} + W_{ij3}\sigma_j) + 3(\beta - 4Z_{i33}\sigma_i)P_3^2 + 5\gamma P_3^4 + 7\delta P_3^6]\Pi_3 - g_{33kl}\frac{\partial^2 d_{33}}{\partial x_k \partial x_l} =$$
$$\alpha_T P_3 + \beta_T P_3^3 + \gamma_T P_3^5. \quad (A.9c)$$

Here, the parameters $\alpha = \alpha_T(T - T_C^\alpha)$, $\beta = \beta_0 + \beta_T(T - T_C^\beta)$, $\gamma = \gamma_0 + \gamma_T(T - T_C^\gamma)$, and $\delta$ are the 2-nd, 4-th, 6-th and 8-th order Landau expansion coefficients in the $P_3$-powers of the free energy corresponding to the bulk BTO [85, 86]. The values $\sigma_i$ denote diagonal components of a stress tensor in the Voigt notation, and the subscripts $i, j = 1 - 6$. The values $Q_{i3}$, $Z_{i33}$, and $W_{ij3}$ are the components of a linear and two nonlinear electrostriction strain tensors in the Voigt notation, respectively [87]. The values $g_{33kl}$ are polarization gradient coefficients in the matrix notation, and the subscripts $k, l = 1 - 3$. The boundary condition for $P_3$ at the nanoparticle surface S is "natural", i.e., $g_{33kl} n_k \frac{\partial P_3}{\partial x_l}\big|_S = 0$, where $\vec{n}$ is the outer normal to the surface.

The value $E_3$ is an electric field component, which is a superposition of external and depolarization fields, $E_3^0$ and $E_3^e$, respectively. The quasi-static field $E_3$ is related to the electric potential $\varphi$ as $E_3 = -\frac{\partial \varphi}{\partial x_3}$. The potential $\varphi$ satisfies the Poisson equation inside the nanoparticle and the Laplace equation outside the charge-screening shell (see **Appendix A1** for details).

The parameters of BTO used in our calculations are taken from Pertsev et al. [85] and Wang et al.[86]. They are listed in **Tables AI** in **Appendix A1**. The scalar parameters $\alpha, \beta, \gamma$, and $\delta$ in Eq.(A.9) are related with tensorial coefficients, $a_i$, $a_{ij}$, $a_{ijk}$ and $a_{ijkl}$ in the LGD free energy (A.2) in a conventional way. Namely, $\alpha = 2a_1$, $\beta = 4a_{11}$, $\gamma = 6a_{111}$, and $\delta = 8a_{1111}$ in the tetragonal phase of the core.

The EC temperature change $\Delta T_{EC}$, can be calculated from the expression [33]:

$$\Delta T_{EC} = -T \int_{E_1}^{E_2} \frac{1}{\rho_P C_P}\left(\frac{\partial P}{\partial T}\right)_E dE \cong T \int_{E_1}^{E_2} \frac{1}{\rho_P C_P} \Pi_3 \, dE, \quad (A.10a)$$

where $\rho_P$ is the volume density, $T$ is the ambient temperature, and $C_P$ is the specific heat of the nanoparticle. The entropy and heat variations $\Delta S$ and $\Delta Q$, induced by the EC effect, are given by the expression, $\Delta S = \frac{\Delta Q}{T} = -\int_{E_1}^{E_2} \left(\frac{\partial P}{\partial T}\right)_E dE$, and thus $\left(\frac{\partial S}{\partial E}\right)_T = -\left(\frac{\partial P}{\partial T}\right)_E$ and $\left(\frac{\partial Q}{\partial E}\right)_T = -T\left(\frac{\partial P}{\partial T}\right)_E$.

For ferroics, the specific heat depends on the electric polarization (and so on the external field) and can be modeled as follows [88]:

$$C_P = C_P^0 - T\frac{\partial^2 g}{\partial T^2}, \qquad (A.10b)$$

where $C_P^0$ is the polarization-independent part of specific heat, and $g$ is the density of the LGD free energy (A.2). According to the experiment, the specific heat usually has a maximum in the first order ferroelectric phase transition point, which height is about 10 – 30 % of the $C_p$ value near $T_C$ (see, e.g. [89]). The mass density and the polarization-independent part of the specific heat of BTO are $\rho_P = 6.02 \cdot 10^3$ kg/m³ and $C_P^0 = 4.6 \cdot 10^2$ J/(kg K), respectively.

In the case of natural boundary conditions used here, $g_{33ij} n_i \frac{\partial P_3}{\partial x_j} = 0$, small screening length $\lambda \leq 0.5$ nm and relatively high gradient coefficients of polarization, $|g_{ijkl}| > 10^{-11}$ C⁻²m³J, polarization gradient effects can be neglected in a single-domain state, which is revealed to have minimal energy compared to polydomain states.

**A3. Models of effective media**

The Landau approximation of a liner mixture [48], which is valid for $v_P \ll 0.1$, gives the expression for $\varepsilon_{eff}$:

$$\varepsilon_{eff}^L \approx \varepsilon_P v_P + \varepsilon_M(1 - v_p). \qquad (A.11a)$$

The Maxwell-Garnett model [49], which is suitable for spherical nanoparticles, provided their volume fraction is small ($v_P$ less than 10%), gives the expression for $\varepsilon_{eff}$:

$$\varepsilon_{eff}^{MG} = \varepsilon_M \left(1 + \frac{3v_P(\varepsilon_P - \varepsilon_M)}{(\varepsilon_P + 2\varepsilon_M) - v_P(\varepsilon_P - \varepsilon_M)}\right), \qquad (A.11b)$$

The Bruggeman model [50], which is applicable if the volume fraction of nanoparticles is less (20-30)%, gives the expression for $\varepsilon_{eff}$

$$\varepsilon_{eff}^B = \frac{1}{4}\left(H_b + \sqrt{8\varepsilon_M \varepsilon_P + H_b^2}\right), \qquad (A.11c)$$

where $H_b = [3(v_P - 1) - 1]\varepsilon_M + (3v_P - 1)\varepsilon_P$. The Lichtenecker-Rother model of a logarithmic mixture [51], which is valid for a random distribution of shapes and orientations of each component, yields the equation for the $\varepsilon_{eff}$:

$$\log \varepsilon_{eff}^{LR} \approx v_P \log \varepsilon_P + (1 - v_P) \log \varepsilon_M. \qquad (A.11d)$$

The dependences of $\varepsilon_{eff}$ on $v_P$ calculated from Eqs.(A.11) for core permittivity $\varepsilon_c = 350$ and matrix permittivity $\varepsilon_M = 30$, which is possible to reach for poly-vinylidene-fluoride (**PVDF**), $\varepsilon_M = 3$, which is close to the static permittivity of poly-vinyl-butyral (**PVB**) and ethyl-cellulose (**ETC**), or $\varepsilon_M = 1$ for air, are shown in **Fig. A1(a), A1(b)** and **A1(c)**, respectively. As one can see, the Landau model corresponds to the maximal $\varepsilon_{eff}$, and the Maxwell-Garnett model corresponds to the minimal $\varepsilon_{eff}$, and both models are not applicable for $v_P > 0.1$. The Lichtenecker-Rother model of effective media, which gives the median results, enables the charge density at any position to be replaced by the mean charge density of the mixture and does not impose severe limitations on $v_P$. According to the Lichtenecker-Rother model, the nanoparticle fraction $v_P$ must exceed 0.25 to increase the ratio $\varepsilon_{eff}/\varepsilon_c$ above 0.1 for the PVDF matrix, $v_P$ should be higher than 0.5 to reach $\varepsilon_{eff}/\varepsilon_c > 0.1$ in the PVB matrix, and $v_P > 0.55$ is required for $\varepsilon_{eff}/\varepsilon_c > 0.1$ for air or gaseous matrix.

To the best of our knowledge, the Landau model of the linear mixture and the Lichtenecker-Rother model of the logarithmic mixture can be generalized for more than two components:

$$\varepsilon_{eff}^L \approx \sum_{i=1}^{N} \varepsilon_{Pi} v_{Pi} + \varepsilon_M (1 - \sum_{i=1}^{N} v_{Pi}), \quad \varepsilon_{eff}^{LR} \approx \prod_{i=1}^{N} \varepsilon_{Pi}^{v_{Pi}} \varepsilon_M^{1-\sum_{j=1}^{N} v_{Pj}}, \quad (A.12)$$

where $N$ is the number of components. However, the fractions of ferroelectric inclusions should be small for the Landau model applicability, while these limitations are absent for the Lichtenecker-Rother model. Therefore, the Lichtenecker-Rother model seems the most suitable for further use because we need to consider dense composites with $v_P \geq 0.2$.

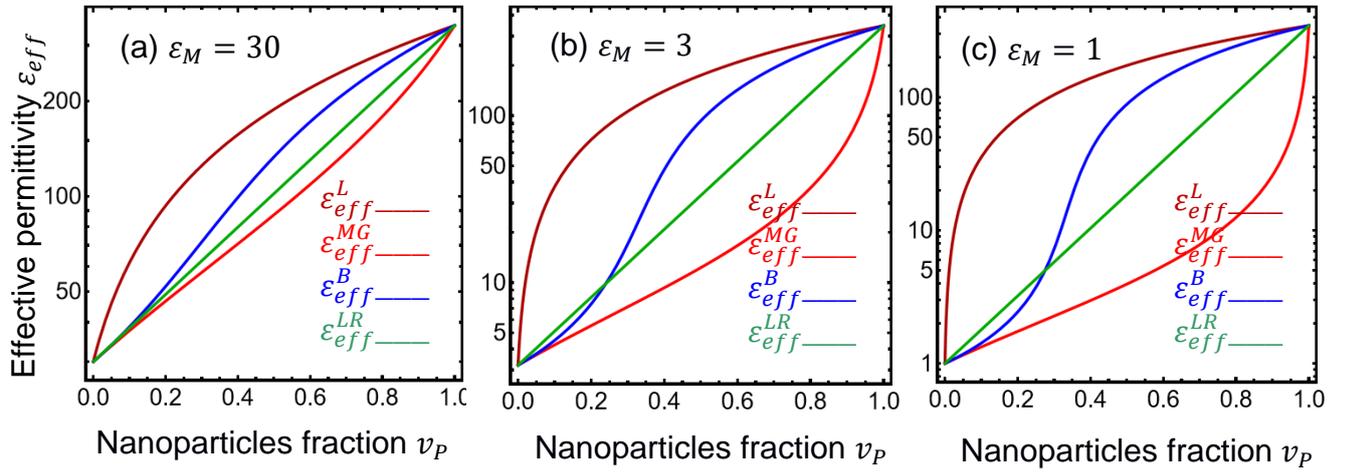

**FIGURE A1.** The dependence of the effective dielectric permittivity of the composite $\varepsilon_{eff}$ on the particle fraction $v_P$ calculated using Eqs.(9) for the static permittivity $\varepsilon_c = 350$ and $\varepsilon_M = 30$ **(a)**, 3 **(b)** and 1 **(c)**.

# APPENDIX B. Characterization of BaTiO₃ nanopowder, nanocomposite preparation details and electrophysical measurements

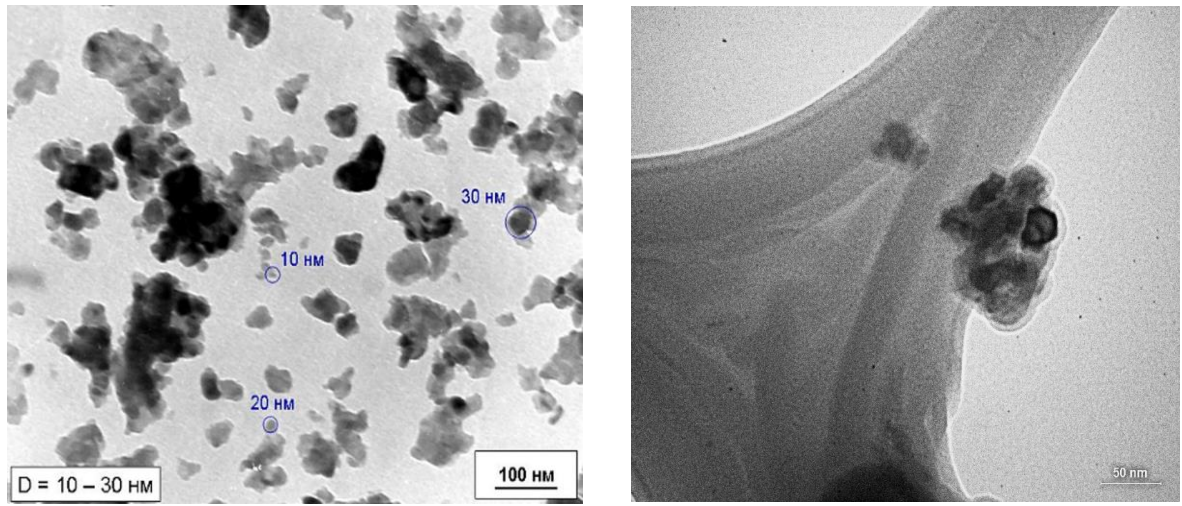

**FIGURE B1.** The structure of the BaTiO₃ nanopowder (TEM images).

**Table BI.** Chemical composition of BaTiO₃ nanopowder (chemical analyzer EXPERT 3L)

| Element | Content (%) | Error (%) |
|---|---|---|
| Ba | 81.710 | 0.221 |
| Ti | 15.900 | 0.196 |
| V  | 1.579  | 0.126 |
| Mn | 0.364  | 0.031 |
| Sr | 0.355  | 0.005 |
| Sn | 0.080  | 0.007 |
| Zn | 0.012  | 0.002 |

**Table BII.** The capacitance, conductivity, and tangent of dielectric losses angle of the composite films "BTO nanoparticles + PVB" (samples TCS-27 and TCS-34)

| Area S (in mm²) | Capacitance $C_f$ (in pF) | Conductivity σ (in μS) | Tangent of the dielectric losses angle tgδ |
|---|---|---|---|
| Sample TCS-27, thickness $h = 18$ μm | | | |
| 1 | 6.1, 6.0, 6.3, 5.98, 5.98, 5.7, 5.23 | 2.13, 2.0, 1.96, 1.81, 1.91, 1.82, 1.68 | 0.0549, 0.054, 0.049, 0.0490, 0.0542, 0.0517, 0.0507 |
| 2.25 | 11.2, 11.54, 11.83 | 2.7, 2.66, 2.69 | 0.0383, 0.0372, 0.0363 |
| 4 | 16.56, 20.0, 13.9, 19.76, 26.72, 26.72 | 3.4, 5.8, 3.2, 4.4, 4.9, 4.9 | 0.0343, 0.0468, 0.0360, 0.347, 0.0299, 0.0299 |
| 6.25 | 26.72, 30.0, 30.13 | 4.9, 5.9, 6.5 | 0.0299, 0.0314, 0.0348 |
| 9 | 36.32, 41.8, 42.74, 42.0 | 6.2, 7.4, 7.5, 7.5 | 0.0273, 0.0287, 0.0287, 0.0280 |
| Sample TCS-34, thickness $h = 8.5$ μm | | | |
| 1 | 13.98, 14.1, 15.24, 15.13, 15.73, 15.66, 16.2, 14.33, 15.7, 16.1, 15.64 | 5.2, 4.6, 5.1, 4.52, 3.1, 3.2, 4.2, 3, 3.1, 3.5, 3.9 | 0.0603, 0.0503, 0.0534, 0.0483, 0.033, 0.034, 0.0407, 0.033, 0.032, 0.0346, 0.0407 |
| 2.25 | 30, 33.12, 31.6, 31.7, 35.46, 31 | 5.3, 5.9, 6, 5.4, 6.3, 6.7 | 0.028, 0.0289, 0.031, 0.0275, 0.0285, 0.0309 |
| 4 | 68.5, 65.4, 58.89, 57.67, 57.3 | 11.1, 10.8, 11.2, 9.3, 9.6 | 0.026, 0.0266, 0.0304, 0.026, 0.0274 |
| 6.25 | 97.7, 83.77, 86.2 | 15.7, 13, 13.3 | 0.0257, 0.0251, 0.0245 |
| 9 | 71.6, 114 | 12, 17.6 | 0.0275, 0.024 |

Shown in **Fig. B2** are current-voltage (I-V) characteristics for three chosen samples. One can see that the I-V characteristics manifest a clear ohmic behavior with negligible traces of loop-wise run in the backward branches. Therefore, the studied composite samples in the steady-state regime of the vertical electric transport do not reveal any noticeable nonlinearities for the voltages less than 10 V. Since the thermodynamic coercive field of BTO nanoparticles is about 20 mV/nm (see **Figs. 1(b)-(d)**), and the maximal field does not exceed 0.56 mV/nm for TCS-27 and 1.18 mV/nm for TCS-34 in our experiments, the voltage amplitude 10 V is likely too small for the of any observation ferroelectric nonlinearities. However, the experiment's goal was not the polarization switching but the low-voltage linear regime to determine the effective media parameters.

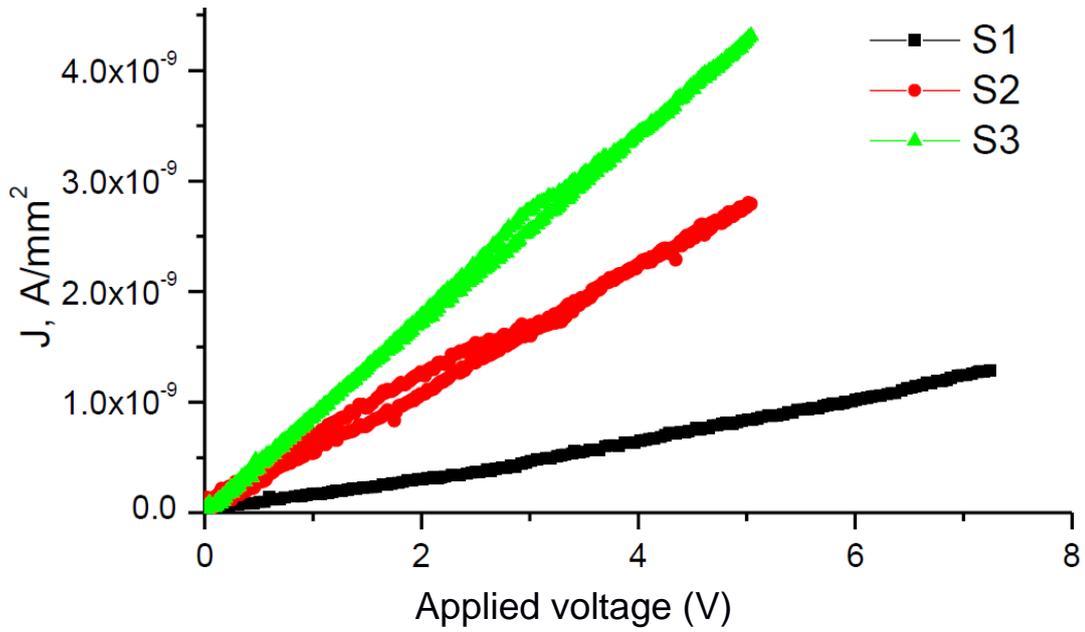

**FIGURE B2.** Current-voltage characteristics for the sample TCS-27 with different contact areas $S1 = 9$ mm$^2$, $S2 = 6.25$ mm$^2$, and $S3 = 4$ mm$^2$.

Shown in **Fig. B3(a)** and **B3(b)** are waveforms of the charge-voltage (Q-V) characteristics measured both with the triangular and sinusoidal voltage scanning at the 1 kHz frequency. As seen from **Fig. B3(c)**, the electric breakdown is observed at sufficiently high voltage magnitude, which character should be investigated in more detail. Also, all Q-V characteristics have an almost elliptic shape for a triangular-shaped periodic voltage and purely elliptic shape for a sinusoidal shape. This means that the external electric field was too weak in the BTO nanoparticles, which were effectively screened from the field since $\varepsilon_{eff}/\varepsilon_c < 0.1$. Therefore, measuring any reliable polar characteristic related to the spontaneous polarization reversal in the BTO nanoparticles was impossible. We observed multiple electric breakdowns during the Q-V and I-V scanning measurements, which may be associated with the appearance of silver bridges under the vertical silver pasted electrodes.

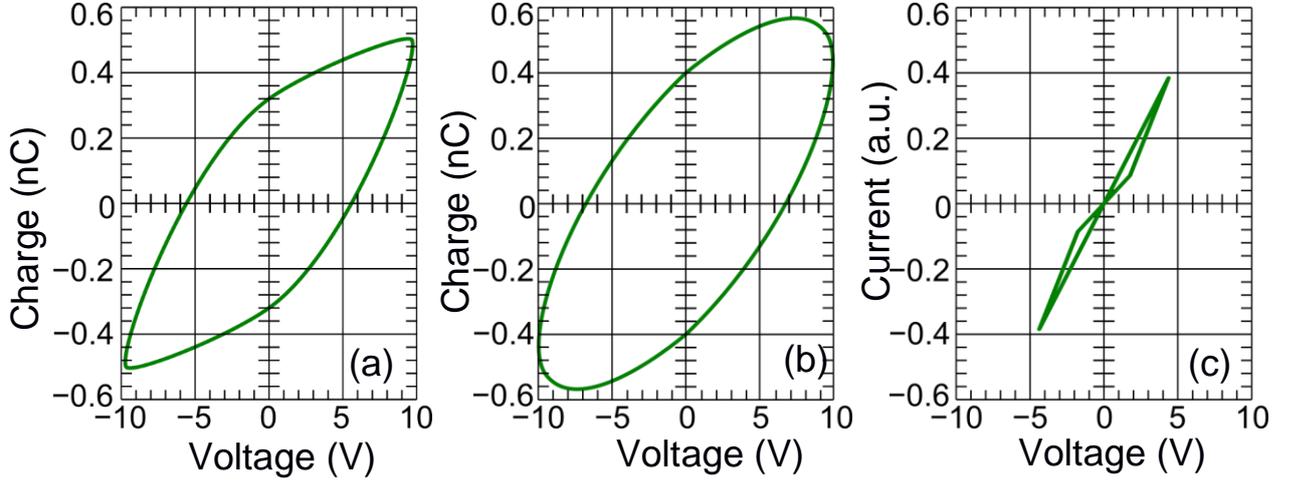

**FIGURE B3.** Typical oscillograms of Q-V curves of the samples TCS-27 and TCS-34 measured for triangular-shape **(a)** and sinusoidal-shape **(b)** periodic voltages. **(c)** The moment of electric breakdown for I-V curves.

The breakdown effect, shown in **Fig. B3(c)**, can be corroborated by the FEM results. Two BTO nanoparticles surrounded by four ellipsoidal semiconducting ($WS_2$) or conducting (Ag) nanoparticles, which are placed near the substrate-electrode, are shown in **Fig. B4(a)**. Distributions of electric potential $\phi$, normal and lateral components of the quasi-static electric field, the $E_3$ and $E_1$ are shown in **Fig. B4(b)-(h)**, respectively. We put $\varepsilon_M = 3$ and $\nu_p = 0.3$ to reach $\varepsilon_{eff} \approx 10$ in accordance with the experimental results presented in **Table BII**. Also, we use the following parameters of semiconducting nanoparticles: semi-axes are 10 nm and 5 nm, respectively, Debye-Hukkel screening radius $R_D$ is 2 nm inside the $WS_2$ nanoparticles and 0.2 nm inside the Ag nanoparticles. Like in **Fig. 3**, we do not show the depolarization field induced by the spontaneous polarization in the nanoparticles in **Fig. B4**, because we aim to compare the external field components with the coercive field.

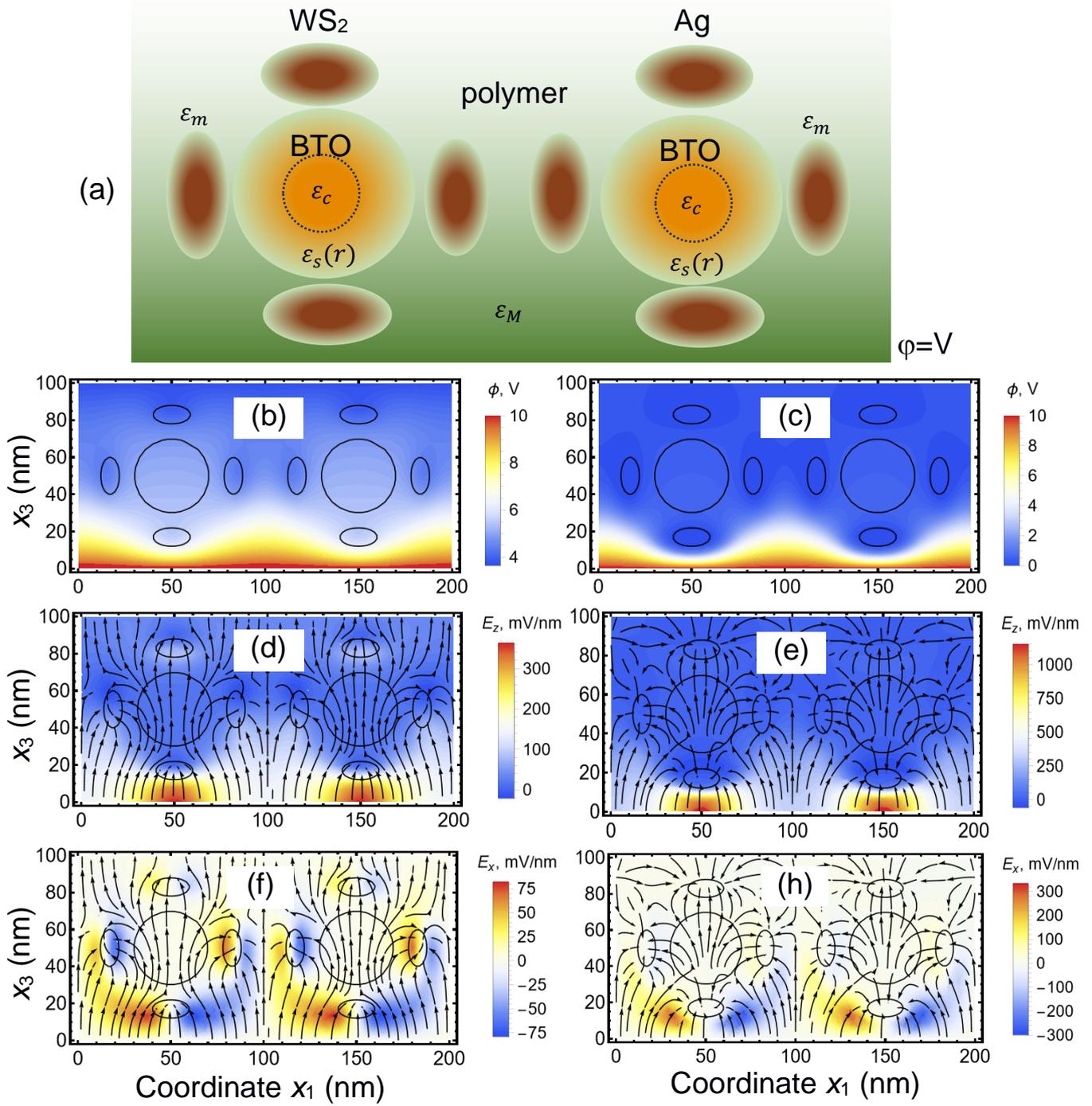

**FIGURE B4.** (a) Schematics of two BTO nanoparticles surrounded by four ellipsoidal semiconducting (WS$_2$) or conducting (Ag) nanoparticles, which are placed near the substrate-electrode. Distributions of electric potential $\phi$ (b, c), normal (d, e), and lateral (f, h) components of the quasi-static electric field in the system. Parameters: $V = 10$ V is applied to a 200-nm thick capacitor; $\varepsilon_c = 350$, $\varepsilon_M = 10$; $R_D = 2$ nm and $\varepsilon_m = 16$ (in the left column for WS$_2$); $\varepsilon_m \geq 100$ and $R_D = 0.2$ nm (in the right column for Ag). The radius of BTO nanoparticles is 20 nm, and their diffuse shell is 5 nm thick; the ellipsoidal nanoparticles semi-axes are 10 nm and 5 nm, respectively.

As expected, the semiconducting WS$_2$ nanoparticles, and especially metallic Ag nanoparticles of high concentration, effectively screen the BTO nanoparticles from external electric fields. For instance, $E_3$ varies from +19 mV/nm to -22 mV/nm inside the BTO nanoparticle surrounded by Ag nanoparticles. Since the field is higher than the coercive field $E_c$ for a single-domain BTO nanoparticle ($E_c \approx 20$ mV/nm, see **Figs. 1(b)-(d)**), Ag nanoparticles, which induce a strongly inhomogeneous electric field inside the BTO nanoparticles, can lead to domain formation in them.

The very strong enhancement of the electric field outside the WS$_2$ and/or Ag nanoparticles exists near the electrode with potential $\varphi = V$. Namely, $E_3$ reaches 350 mV/nm between the electrode and the WS$_2$ nanoparticle, and $E_3$ is more than 1 V/nm between the electrode and the Ag nanoparticle. Since the local field enhancement outside the Ag nanoparticles can be very high (> 1 V/nm), the field leads to the electric breakdown, which looks like the conductive "bridges" between the Ag nanoparticles in the polymer, and this explains the experimental observations shown in **Fig. B3(c)**.

**APPENDIX C. Calculation details of the electrocaloric response of ferroelectric composites**

Since the application of the electric field during the EC heating and cooling are adiabatic processes, the EC response of a ferroelectric composite consisting of the BTO nanoparticles incorporated in a heat conductive media can be estimated from the heat-balance equation:

$$(\Delta T_{P-M} + T)(v_P \rho_P C_P + v_M \rho_M C_M) = v_P \rho_P C_P (\Delta T_{EC} + T) + v_M \rho_M C_M T. \qquad (C.1)$$

Here $\Delta T_{P-M}$ is the temperature change of the composite induced by the EC effect in the BTO nanoparticles, $v_P$ and $v_M = 1 - v_P$ are the relative volumes of the nanoparticles and the media, respectively, $\rho_P$ and $\rho_M$ are their volume densities, $C_P$ and $C_M$ are their specific heat capacitances, and $T$ is the initial temperature of the media. The solution of Eq.(12a) is:

$$\Delta T_{P-M} = \frac{v_P \rho_P C_P}{v_P \rho_P C_P + v_M \rho_M C_M} \Delta T_{EC}. \qquad (C.2)$$

As one can see from (C.2), $|\Delta T_{P-M}| \leq |\Delta T_{EC}|$, and thus, it makes sense to prepare very dense composites with $v_P \geq v_M$ and/or to use polymers and/or fillers with lower heat mass $\rho_M C_M \leq \rho_P C_P$. Below, we analyze the numerical estimates of $\Delta T_{P-M}$ for dense composites, whose experimental results were presented in the previous section.

The mass density and heat capacity of BTO are $\rho_P = 6.02 \cdot 10^3$ kg/m$^3$ and $C_P = 4.6 \cdot 10^2$ J/(kg K), respectively. The product of these values, $\rho_P C_P$, is significantly higher than the corresponding products for such polymers as PVB ($\rho_{PVB} = 1.1 \cdot 10^3$ kg/m$^3$, $C_{PVB} = 1.75 \cdot 10^2$ J/(kg K)), PVDF ($\rho_{PVDF} = 1.78 \cdot 10^3$ kg/m$^3$, $C_{PVDF} = 1.2 \cdot 10^2$ J/(kg K)) and ETC ($\rho_{ET} = 1.14 \cdot$ $10^3$ kg/m³, $C_{ET} = 2.0 \cdot 10^2$ J/(kg K)), some of which are used in our experiments. Therefore, the dependence of $\Delta T_{P-M}$ on the BTO particle fraction $v_P$ and their EC cooling temperature $\Delta T_{EC}$, shown in **Fig. C1(a)** and **C1(c)**, is almost the same for all three polymers, PVB, PVDF, and ETC. To achieve a significant EC cooling, $\Delta T_{P-M} < -5$ K, the fraction $v_P$ should be higher than 0.25 for the average core radius 6 nm and higher than 0.85 for the average core radius 60 nm.

For the BTO nanopowders in the air, the product $\rho_{Air} C_{Air}$ is about $10^3$ times smaller than the product $\rho_P C_P$, because $\rho_{Air} = 1.3$ kg/m³ and $C_{Air} = 700$ J/(kg K). Therefore, the dependence of $\Delta T_{P-M}$ on $v_P$ and $\Delta T_{EC}$, calculated for the BTO nanopowders and shown in **Fig. C1(b)** and **C1(d)**, is almost $v_P$-independent and looks very different from **Fig. C1(a)** and **C1(c)**, because $\Delta T_{P-M} \approx \Delta T_{EC}$ for $0.001 < v_P \leq 1$.

It may seem from the comparison of the left and right columns in **Fig. C1** that it is reasonable to place the ferroelectric nanoparticles in the air (or gas with a very low product $\rho_M C_M$) to reach the equality $\Delta T_{P-M} \approx \Delta T_{EC}$ in the thermodynamical equilibrium state. Let us underline that the simple expression (C.2) has a practical sense only if the heat flux from ferroelectric nanoparticles can be transferred enough rapidly to the effective media and then from the media to the working substance, which should be cooled or heated by the EC effect in the particles. Thus, neither air nor any other media with low heat conductance is not suitable for use as the composite. According to the Newton heat exchange law, the heat flux coming from the EC effect in ferroelectric nanoparticles can be estimated as:

$$\Delta Q_{P-M} = \kappa_{eff} \Delta T_{P-M} \sim \frac{\kappa_{PM} v_P \rho_P C_P}{v_P \rho_P C_P + v_M \rho_M C_M} \Delta T_{EC}, \quad (C.3)$$

where $\kappa_{eff}$ is the effective heat transfer coefficient of the nanocomposite and $\kappa_{PM}$ is the heat transfer coefficient on the nanoparticle–matrix interface. According to Eq.(C.3), one is required to use the pair nanoparticle – matrix with the maximal $\kappa_{PM}$ value to reach the maximal heat flux. Thus, it is reasonable to incorporate fillers with high heat conductance and relatively low heat capacitance (such as metallic nanoparticles), which act as heat sinks and can facilitate the heat transfer from the ferroelectric nanoparticles to the effective media and from the media to the working substance.

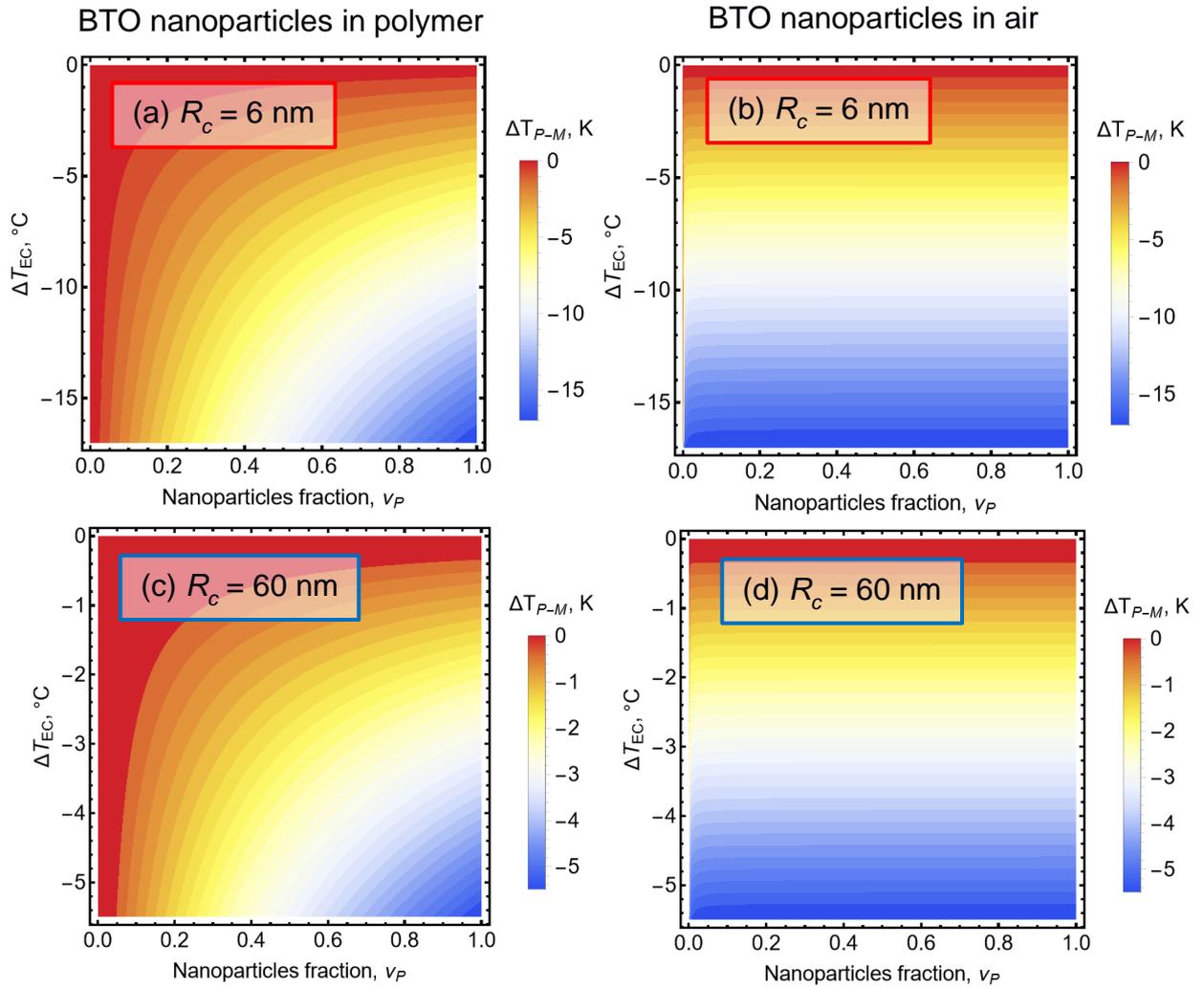

**FIGURE C1.** The dependence of $\Delta T_{P-M}$ on the nanoparticle fraction $v_P$ and EC cooling temperature $\Delta T_{EC}$ calculated for BTO core-shell nanoparticles with the core radius $R_c = 6$ nm **(a, b)** and 60 nm **(c, d)** embedded in different polymers, such as PVB, PVDF, or ETC **(a, c)**, and for BTO nanopowders **(b, d)**.

### References

[1]     Y. Liu, J. F. Scott, B. Dkhil. Direct and indirect measurements on electrocaloric effect: Recent developments and perspectives. Appl. Phys. Rev. **3**, 031102 (2016).

[2]     E. Fatuzzo, H. Kiess, and R. Nitsche. Theoretical efficiency of pyroelectric power converters. J. Appl. Phys. **37**, 510 (1966)

[3]     P. D. Thacher. Electrocaloric effects in some ferroelectric and antiferroelectric Pb(Zr, Ti)O$_3$ compounds. J. Appl. Phys. **39**, 1996 (1968)

[4]     A. S. Mischenko, Q. Zhang, J. F. Scott, R.W. Whatmore, N.D. Mathur. "Giant electrocaloric effect in thin-film PbZr$_{0.95}$Ti$_{0.05}$O$_3$". Science **311**, 1270 (2006).

[5]     R. Kumar, S. Singh, Giant electrocaloric and energy storage performance of [(K$_{0.5}$Na$_{0.5}$)NbO$_3$](1−x) − [LiSbO$_3$]x nanocrystalline ceramics. Sci Rep. **8**, 3186 (2018).

[6] J. F. Scott. Electrocaloric materials. Annual Review of Materials Research, **41**, 229 (2011)

[7] M.C. Rose, R. E. Cohen. Giant electrocaloric effect around TC. Phys. Rev. Lett. **109**, 187604 (2012).

[8] X. Moya, S. Kar-Narayan, N. D. Mathur. Caloric materials near ferroic phase transitions. Nat. Mat. **13**, 439 (2014).

[9] D. Caruntu and T. Rostamzadeh and T. Costanzo and S. S. Parizi and G. Caruntu. Solvothermal synthesis and controlled self-assembly of monodisperse titanium-based perovskite colloidal nanocrystals, Nanoscale **7**, 12955 (2015).

[10] S. Starzonek, S. J. Rzoska, A. Drozd-Rzoska, K. Czupryński, and S. Kralj, Impact of ferroelectric and superparaelectric nanoparticles on phase transitions and dynamics in nematic liquid crystals, Phys. Rev. E **96**, 022705 (2017).

[11] Z. Hanani, D. Mezzane, M. Amjoud, M. Lahcini, M. Spreitzer, D. Vengust, A. Jamali, M. El Marssi, Z. Kutnjak, and M. Gouné. "The paradigm of the filler's dielectric permittivity and aspect ratio in high-k polymer nanocomposites for energy storage applications." Journal of Materials Chemistry **C 10**, 10823 (2022) https://doi.org/10.1039/d2tc00251e

[12] M. Kumari, M. Chahar, S. Shankar, and O. P. Thakur. "Temperature dependent dielectric, ferroelectric and energy storage properties in Bi0.5Na0.5TiO3 (BNT) nanoparticles." Materials Today: Proceedings **67**, 688 (2022), https://doi.org/10.1016/j.matpr.2022.06.542

[13] Z. Luo, Z. Ye, B. Duan, G. Li, K. Li, Z. Yang, S. Nie, T. Chen, L. Zhou, and P. Zhai. "SiC@ BaTiO3 core-shell fillers improved high temperature energy storage density of P (VDF-HFP) based nanocomposites." Composites Science and Technology **229**, 109658 (2022), https://doi.org/10.1016/j.compscitech.2022.109658

[14] Z. Fan, S. Gao, Y. Chang, D. Wang, X. Zhang, H. Huang, Y. He, and Q. Zhang. "Ultra-superior high-temperature energy storage properties in polymer nanocomposites via rational design of core–shell structured inorganic antiferroelectric fillers." Journal of Materials Chemistry A **11**, 7227 (2023), https://doi.org/10.1039/D2TA09658G

[15] G. Cook, J. L. Barnes, S. A. Basun, D. R. Evans, R. F. Ziolo, A. Ponce, V. Yu. Reshetnyak, A. Glushchenko, and P. P. Banerjee, Harvesting single ferroelectric domain stressed nanoparticles for optical and ferroic applications. J. Appl. Phys. **108**, 064309 (2010); https://doi.org/10.1063/1.3477163

[16] G. Zhang, X. Zhang, T. Yang, Qi Li, L.-Q. Chen, S. Jiang, and Q. Wang. Colossal room-temperature electrocaloric effect in ferroelectric polymer nanocomposites using nanostructured barium strontium titanates. ACS nano **9**, 7164 (2015). https://doi.org/10.1021/acsnano.5b03371

[17] E. Erdem, H.-Ch. Semmelhack, R. Bottcher, H. Rumpf, J. Banys, A.Matthes, H.-J. Glasel, D. Hirsch, E. Hartmann. Study of the tetragonal-to-cubic phase transition in PbTiO$_3$ nanopowders. J. Phys.: Condens. Matter. **18**, 3861 (2006).

[18] J. Zhu, W. Han, H. Zhang, Z. Yuan, X. Wang, L. Li, and C. Jin. Phase coexistence evolution of nano BaTiO$_3$ as function of particle sizes and temperatures. J. Appl. Phys. **112**, 064110 (2012), https://doi.org/10.1063/1.4751332

[19]     S. A. Basun, G. Cook, V. Y. Reshetnyak, A. V. Glushchenko, and D. R. Evans, Dipole moment and spontaneous polarization of ferroelectric nanoparticles in a nonpolar fluid suspension. Phys. Rev. B **84**, 024105 (2011), https://doi.org/10.1103/PhysRevB.84.024105

[20]     D. R. Evans, S. A. Basun, G. Cook, I. P. Pinkevych, and V. Yu. Reshetnyak. Electric field interactions and aggregation dynamics of ferroelectric nanoparticles in isotropic fluid suspensions. Phys. Rev. B, **84,** 174111 (2011), https://doi.org/10.1103/PhysRevB.84.174111

[21]     A. Lorenz, N. Zimmermann, S. Kumar, D. R. Evans, G. Cook, and H.-S. Kitzerow. Doping the nematic liquid crystal 5CB with milled $BaTiO_3$ nanoparticles. Phys. Rev. E **86**, 051704 (2012).

[22]     U. Idehenre, Y. A. Barnakov, S. A. Basun, and D. R. Evans. Spectroscopic studies of the effects of mechanochemical synthesis on $BaTiO_3$ nanocolloids prepared using high-energy ball-milling. J. Appl. Phys. **124**, 165501 (2018)

[23]     D. Yadlovker, S. Berger. Uniform orientation and size of ferroelectric domains. Phys. Rev. B **71**, 184112 (2005).

[24]     D. Yadlovker, S. Berger. Reversible electric field induced nonferroelectric to ferroelectric phase transition in single crystal nanorods of potassium nitrate. Appl. Phys. Lett. **91**, 173104 (2007).

[25]     Schilling, A., R. M. Bowman, G. Catalan, J. F. Scott, and J. M. Gregg. "Morphological control of polar orientation in single-crystal ferroelectric nanowires." Nano letters **7**, 3787 (2007), https://doi.org/10.1021/nl072260l

[26]     H. Huang, G. Zhang, X. Ma, D. Liang, J. Wang, Y. Liu, Q. Wang, and L.-Q. Chen. "Size effects of electrocaloric cooling in ferroelectric nanowires." Journal of the American Ceramic Society **101**, 1566 (2018), https://doi.org/10.1111/jace.15304

[27]     L. Chen, Y. Li, C. Li, H. Wang, Z. Han, He Ma, G. Yuan et al. "Thickness dependence of domain size in 2D ferroelectric $CuInP_2S_6$ nanoflakes." AIP Advances **9**, no. 11 (2019). https://doi.org/10.1063/1.5123366

[28]     M. M. Kržmanc, H. Uršič, A. Meden, R. C. Korošec, and D. Suvorov. "$Ba_{1-x}Sr_xTiO_3$ plates: synthesis through topochemical conversion, piezoelectric and ferroelectric characteristics." Ceramics International **44**, 21406 (2018). https://doi.org/10.1016/j.ceramint.2018.08.198

[29]     M. M. Kržmanc, B. Jančar, H. Uršič, M. Tramšek, and D. Suvorov. "Tailoring the shape, size, crystal structure, and preferential growth orientation of $BaTiO_3$ plates synthesized through a topochemical conversion process." Crystal Growth & Design **17**, 3210 (2017), https://doi.org/10.1021/acs.cgd.7b00164

[30]     Y. A. Barnakov, I. U. Idehenre, S. A. Basun, T. A. Tyson, and D. R. Evans. Uncovering the mystery of ferroelectricity in zero dimensional nanoparticles. Nanoscale Advances **1**, 664 (2019)

[31]     H. Zhang, S. Liu, S. Ghose, B. Ravel, I. U. Idehenre, Y. A. Barnakov, S. A. Basun, D. R. Evans, and T. A. Tyson. Structural Origin of Recovered Ferroelectricity in $BaTiO_3$ Nanoparticles. Phys. Rev. B **108**, 064106 (2023), https://doi.org/10.1103/PhysRevB.108.064106

[32]     A. N. Morozovska, E. A. Eliseev, S. V. Kalinin, and D. R. Evans. Strain-Induced Polarization Enhancement in $BaTiO_3$ Core-Shell Nanoparticles. https://doi.org/10.48550/arXiv.2308.11044.

[33]     A. N. Morozovska, E. A. Eliseev, M. D. Glinchuk, H. V. Shevliakova, G. S. Svechnikov, M. V. Silibin, A. V. Sysa, A. D. Yaremkevich, N. V. Morozovsky, and V. V. Shvartsman. Analytical description

of the size effect on pyroelectric and electrocaloric properties of ferroelectric nanoparticles. Phys.Rev. Materials **3**, 104414 (2019) https://link.aps.org/doi/10.1103/PhysRevMaterials.3.104414

[34]  P. Perriat, J. C. Niepce, G. Caboche. Thermodynamic considerations of the grain size dependency of material properties: a new approach to explain the variation of the dielectric permittivity of $BaTiO_3$ with grain size. J. Therm. Anal. Calorim. **41**, 635 (1994).

[35]  H. Huang, C. Q. Sun, Z. Tianshu, P. Hing. Grain-size effect on ferroelectric $Pb(Zr_{1-x}Ti_x)O_3$ solid solutions induced by surface bond contraction. Phys. Rev. B **63**, 184112 (2001).

[36]  W. Ma. Surface tension and Curie temperature in ferroelectric nanowires and nanodots. Appl. Phys. A **96**, 915 (2009).

[37]  A. N. Morozovska, M. D. Glinchuk, E.A. Eliseev. Phase transitions induced by confinement of ferroic nanoparticles. Phys. Rev. B **76**, 014102 (2007).

[38]  A. N. Morozovska, Y. M. Fomichov, P. Maksymovych, Y. M. Vysochanskii, and E. A. Eliseev. Analytical description of domain morphology and phase diagrams of ferroelectric nanoparticles. Acta Materialia **160**, 109 (2018).

[39]  E.A. Eliseev, A.N. Morozovska, M.D. Glinchuk, R. Blinc. Spontaneous flexoelectric/flexomagnetic effect in nanoferroics. Phys. Rev. B. **79**, 165433 (2009).

[40]  J.J. Wang, X.Q. Ma, Q. Li, J. Britson, L.-Q. Chen. Phase transitions and domain structures of ferroelectric nanoparticles: Phase field model incorporating strong elastic and dielectric inhomogeneity. Acta Mater. **61**, 7591 (2013).

[41]  J. J. Wang, E. A. Eliseev, X. Q. Ma, P. P. Wu, A. N. Morozovska, and Long-Qing Chen. Strain effect on phase transitions of $BaTiO_3$ nanowires. Acta Mater. **59**, 7189 (2011).

[42]  A.N. Morozovska, I.S. Golovina, S.V. Lemishko, A.A. Andriiko, S.A. Khainakov, E.A. Eliseev. Effect of Vegard strains on the extrinsic size effects in ferroelectric nanoparticles. Phys. Rev. B **90**, 214103 (2014)

[43]  A. N. Morozovska, E. A. Eliseev, G. S. Svechnikov, S. V. Kalinin. Pyroelectric response of ferroelectric nanowires: Size effect and electric energy harvesting. J. Appl. Phys. **108**, 042009 (2010).

[44]  M. Liu, J. Wang. Giant electrocaloric effect in ferroelectric nanotubes near room temperature. Sci. Rep. **5**, Art. No. 7728 (2015).

[45]  C. Xiqu, and C. Fang. Study of electrocaloric effect in barium titanate nanoparticle with core–shell model. Physica B: Cond. Matt. **415**, 14 (2013)

[46]  M. V. Silibin, A. V. Solnyshkin, D. A. Kiselev, A. N. Morozovska, E. A. Eliseev, S. Gavrilov, M. D. Malinkovich, D. C. Lupascu, V. V. Shvartsman. Local ferroelectric properties in PVDF/BPZT nanocomposites: interface effect. J. Appl. Phys. **114**, 144102 (2013)

[47]  M. Silibin, J. Belovickis, S. Svirskas; M. Ivanov, J. Banys; A. Solnyshkin, S. Gavrilov, O. Varenyk, A. Pusenkova, N. Morozovsky, V. Shvartsman, A. Morozovska. Polarization reversal in organic-inorganic ferroelectric composites: modeling and experiment. Applied Physics Letters **107**, 142907 (2015); https://doi.org/10.1063/1.4932661

[48]  L. D. Landau, L. P. Pitaevskii, E. M. Lifshitz *Electrodynamics of continuous media*. Vol. 8. Elsevier (2013). (Translated by J. S. Bell, M. J. Kearsley, and J. B. Sykes).

[49]   J. C. M. Garnett, "Colours in metal glasses and in metallic films," Philos. Trans. R. Soc. London. Ser. A, Contain. Pap. a Math. or Phys. Character, vol. 203, no. 359–371, pp. 385–420 (1904), https://doi.org/10.1098/rsta.1904.0024

[50]   D. A. G. Bruggeman, "Berechnung verschiedener physikalischer Konstanten von heterogenen Substanzen. I. Dielektrizitätskonstanten und Leitfähigkeiten der Mischkörper aus isotropen Substanzen," Ann. Phys., vol. 416, no. 7, pp. 636–664 (1935), https://doi.org/10.1002/andp.19354160705 .

[51]   R. Simpkin, Derivation of Lichtenecker's logarithmic mixture formula from Maxwell's equations. IEEE Transactions on Microwave Theory and Techniques **58** (3), 545 (2010), https://doi.org/10.1109/TMTT.2010.2040406

[52]   S. A. Basun, G. Cook, V. Y. Reshetnyak, A. V. Glushchenko, and D. R. Evans, Dipole moment and spontaneous polarization of ferroelectric nanoparticles in a nonpolar fluid suspension. Phys. Rev. B **84**, 024105 (2011).

[53]   D. R. Evans, S. A. Basun, G. Cook, I. P. Pinkevych, and V. Yu. Reshetnyak. Electric field interactions and aggregation dynamics of ferroelectric nanoparticles in isotropic fluid suspensions. Phys. Rev. B, **84,** 174111 (2011)

[54]   A.K. Tagantsev and G. Gerra, Interface-induced phenomena in polarization response of ferroelectric thin films, J. Appl. Phys. **100**, 051607 (2006).

[55]   S. V. Kalinin, Y. Kim, D. Fong, and A. Morozovska. Surface Screening Mechanisms in Ferroelectric Thin Films and its Effect on Polarization Dynamics and Domain Structures. Rep. Prog. Phys. **81,** 036502 (2018). https://doi.org/10.1088/1361-6633/aa915a

[56]   D. A. Freedman, D. Roundy, and T. A. Arias. Elastic effects of vacancies in strontium titanate: Short- and long-range strain fields, elastic dipole tensors, and chemical strain. Phys. Rev. B **80**, 064108 (2009).

[57]   Y. Kim, A. S. Disa, T. E. Babakol, and J. D. Brock. Strain screening by mobile oxygen vacancies in $SrTiO_3$. Appl. Phys. Lett. **96**, 251901 (2010).

[58]   See Supplemental Material at [URL will be inserted by publisher] for a mathematical formulation of the problem, a table of material parameters, a description of methods, and numerical algorithms

[59]   N.A. Pertsev, A.G. Zembilgotov, A. K. Tagantsev, Effect of Mechanical Boundary Conditions on Phase Diagrams of Epitaxial Ferroelectric Thin Films, Phys. Rev. Lett. **80,** 1988 (1998).

[60]   Y. L. Wang, A. K. Tagantsev, D. Damjanovic, N. Setter, V. K. Yarmarkin, A. I. Sokolov, and I. A. Lukyanchuk, Landau thermodynamic potential for $BaTiO_3$, J. Appl. Phys. **101**, 104115 (2007).

[61]   A. N. Morozovska, E. A. Eliseev, M. E. Yelisieiev, Y. M.Vysochanskii, and D. R. Evans. Stress-Induced Transformations of Polarization Switching in $CuInP_2S_6$ Nanoparticles. Phys. Rev. Applied **19,** 054083 (2023), https://doi.org/10.48550/arXiv.2302.07392

[62]   E. A. Eliseev, A. V. Semchenko, Y. M. Fomichov, M. D. Glinchuk, V. V. Sidsky, V. V. Kolos, Yu. M. Pleskachevsky, M. V. Silibin, N. V. Morozovsky, A. N. Morozovska. Surface and finite size effects impact on the phase diagrams, polar and dielectric properties of $(Sr,Bi)Ta_2O_9$ ferroelectric nanoparticles. J. Appl. Phys. **119**, 204104 (2016).

[63]     A. Kvasov and A.K. Tagantsev. Role of high-order electromechanical coupling terms in thermodynamics of ferroelectric thin films. Phys. Rev. B **87**, 184101 (2013).

[64]     H. D. Megaw, Temperature changes in the crystal structure of barium titanium oxide. Trans. Faraday Soc. 42A, 224 (1946); Proc. Roy. Soc. 1S9A, 261 (1947). https://doi.org/10.1098/rspa.1947.0038

[65]     The Mathematica (https://www.wolfram.com/mathematica) notebook, which contain the codes, is available per reasonable request.

[66]     S.E. Ivanchenko, I.O. Dulina, S.O. Umerova, A.G. Nikulin, A.V. Ragulya. Formulation and rheology of tape casting suspensions based on $BaTiO_3$ nanopowders, Springer Proceedings in Physics 167, pp. 193–202 (2015), https://doi.org/10.1007/978-3-319-18543-9_11

[67]     B. L. Funt. Dielectric properties of polyvinyl butyral. Canadian Journal of Chemistry **30**, 84 (1952), https://doi.org/10.1139/v52-013

[68]     P. C. Mehendru; Naresh Kumar; V. P. Arora; N. P. Gupta. Dielectric relaxation studies in polyvinyl butyral. *J. Chem. Phys.* **77**, 4232 (1982), https://doi.org/10.1063/1.444334

[69]     R. M. Omer, E. T. B. Al-Tikrity, R. N. Abed, M. Kadhom, A. H. Jawad, E. Yousif. Electrical Conductivity and Surface Morphology of PVB Films Doped with Different Nanoparticles. Prog. Color Colorants Coat. **15**, 191 (2022)

[70]     https://www.kabusa.com/Dilectric-Constants.pdf

[71]     A. N. Morozovska, E. A. Eliseev, P.S. Sankara Rama Krishnan, A. Tselev, E. Strelkov, A. Borisevich, O. V. Varenyk, N. V. Morozovsky, P. Munroe, S. V. Kalinin and V. Nagarajan. Defect thermodynamics and kinetics in thin strained ferroelectric films: the interplay of possible mechanisms. Phys.Rev.B **89**, 054102 (2014) https://doi.org/10.1103/PhysRevB.89.054102

[72]     L. Curecheriu, S.-B. Balmus, M. T. Buscaglia, V. Buscaglia, A. Ianculescu, and L. Mitoseriu. Grain size-dependent properties of dense nanocrystalline barium titanate ceramics. Journal of the American Ceramic Society **95**, 3912 (2012), https://doi.org/10.1111/j.1551-2916.2012.05409.x

[73]     Yi Zheng, Guang-Xin Ni, Chee-Tat Toh, Chin-Yaw Tan, Kui Yao, B. Özyilmaz. "Graphene field-effect transistors with ferroelectric gating." Phys. Rev. Lett. **105**, 166602 (2010), https://doi.org/10.1103/PhysRevLett.105.166602

[74]     W. Y. Kim, H.-D. Kim, T.-T. Kim, H.-S. Park, K. Lee, H. J. Choi, S. Hoon Lee, Jaehyeon Son, Namkyoo Park, and Bumki Min. Graphene-ferroelectric metadevices for nonvolatile memory and reconfigurable logic-gate operations. Nature communications **7**, Article number: 10429 (2016), https://doi.org/10.1038/ncomms10429

[75]     W. Jie, and J. Hao. Time-dependent transport characteristics of graphene tuned by ferroelectric polarization and interface charge trapping. Nanoscale. **10**, 328 (2017), https://doi.org/10.1039/C7NR06485C

[76]     T. Mishonov and E. Penev. Theory of high temperature superconductivity: a conventional approach. World Scientific (2011), ISBN: 9814343145

[77]     S. G. Lu, B. Rožič, Q. M. Zhang, Z. Kutnjak, Xinyu Li, E. Furman, Lee J. Gorny et al. Organic and inorganic relaxor ferroelectrics with giant electrocaloric effect. Applied Physics Letters **97**, 162904 (2010), https://doi.org/10.1063/1.3501975

[78]     B. Peng, H. Fan, and Qi Zhang. A giant electrocaloric effect in nanoscale antiferroelectric and ferroelectric phases coexisting in a relaxor $Pb_{0.8}Ba_{0.2}ZrO_3$ thin film at room temperature. Advanced Functional Materials **23**, 2987 (2013), https://doi.org/10.1002/adfm.201202525

[79]     A.K. Tagantsev and G. Gerra, Interface-induced phenomena in polarization response of ferroelectric thin films, J. Appl. Phys. **100**, 051607 (2006).

[80]     E. A. Eliseev, Y. M. Fomichov, S. V. Kalinin, Y. M. Vysochanskii, P. Maksymovich and A. N. Morozovska. Labyrinthine domains in ferroelectric nanoparticles: Manifestation of a gradient-induced morphological phase transition. Phys. Rev. B **98**, 054101 (2018).

[81]     J. J. Wang, E. A. Eliseev, X. Q. Ma, P. P. Wu, A. N. Morozovska, and Long-Qing Chen. Strain effect on phase transitions of $BaTiO_3$ nanowires. Acta Materialia **59**, 7189 (2011).

[82]     A. N. Morozovska, E. A. Eliseev, Y. A. Genenko, I. S. Vorotiahin, M. V. Silibin, Ye Cao, Y. Kim, M. D. Glinchuk, and S. V. Kalinin. Flexocoupling impact on the size effects of piezo- response and conductance in mixed-type ferroelectrics-semiconductors under applied pressure. Phys. Rev. B **94,** 174101 (2016).

[83]     L. D. Landau, and I. M. Khalatnikov. On the anomalous absorption of sound near a second order phase transition point. In Dokl. Akad. Nauk SSSR, **96**, 469 (1954).

[84]     J. Hlinka, Mobility of Ferroelectric Domain Walls in Barium Titanate, Ferroelectrics **349,** 49 (2007).

[85]     N.A. Pertsev, A.G. Zembilgotov, A. K. Tagantsev, Effect of Mechanical Boundary Conditions on Phase Diagrams of Epitaxial Ferroelectric Thin Films, Phys. Rev. Lett. **80,** 1988 (1998).

[86]     Y. L. Wang, A. K. Tagantsev, D. Damjanovic, N. Setter, V. K. Yarmarkin, A. I. Sokolov, and I. A. Lukyanchuk, Landau thermodynamic potential for $BaTiO_3$, J. Appl. Phys. **101**, 104115 (2007).

[87]     A. Kvasov and A.K. Tagantsev. Role of high-order electromechanical coupling terms in thermodynamics of ferroelectric thin films. Phys. Rev. B **87**, 184101 (2013).

[88]     A. N. Morozovska, V. V. Khist, M. D. Glinchuk, Venkatraman Gopalan, E. A. Eliseev. Linear antiferrodistortive-antiferromagnetic effect in multiferroics: physical manifestations. Phys. Rev. **B 92**, 054421 (2015)

[89]     F. Jona, and G. Shirane. Ferroelectric Crystals, International Series of Monographs on Solid State Physics. Pergamon press (1962).